%% file: main.tex
\documentclass[journal]{vgtc}                % final (journal style)
\usepackage[dvipsnames]{xcolor}
\usepackage[linesnumbered,lined,commentsnumbered]{algorithm2e}
\usepackage[draft]{hyperref}

\ifpdf%                                % if we use pdflatex
  \pdfoutput=1\relax                   % create PDFs from pdfLaTeX
  \usepackage{graphicx}                % allow us to embed graphics files
  \DeclareGraphicsExtensions{.pdf,.png,.jpg,.jpeg} % for pdflatex we expect .pdf, .png, or .jpg files
\else%                                 % else we use pure latex
  \usepackage{graphicx}                % allow us to embed graphics files
  \DeclareGraphicsExtensions{.eps}     % for pure latex we expect eps files
\fi%

\graphicspath{{figures/}{pictures/}{images/}{./}} % where to search for the images

\usepackage{microtype}                 % use micro-typography (slightly more compact, better to read)
\PassOptionsToPackage{warn}{textcomp}  % to address font issues with \textrightarrow
\usepackage{textcomp}                  % use better special symbols
\usepackage{mathptmx}                  % use matching math font
\usepackage{times}                     % we use Times as the main font
\usepackage{cite}                      % needed to automatically sort the references
\usepackage{tabu}                      % only used for the table example
\usepackage{booktabs}                  % only used for the table example
\usepackage[bottom]{footmisc}
\usepackage{stfloats}
\usepackage{flushend}
\usepackage{enumitem}
\setitemize{noitemsep,topsep=0pt,parsep=0pt,partopsep=0pt}

\usepackage{colortbl} % to color a row
\usepackage{multirow}
\definecolor{green}{RGB}{3, 128, 0}
\definecolor{lightOrange}{RGB}{252, 213, 136}
\definecolor{lightGray}{RGB}{204, 220, 235}
\usepackage{pifont}% http://ctan.org/pkg/pifont
\newcommand{\cmark}{\ding{51}}%
\newcommand{\xmark}{\ding{55}}%

\newcommand{\eg}{\textit{e.g.}}
\newcommand{\ie}{\textit{i.e.}}
\newcommand{\qianwen}[1]{\textcolor{black}{#1}}

\newcommand{\mycircle}{%
  \begingroup\normalfont
  \includegraphics[height=\fontcharht\font`\B]{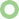}%
  \endgroup
}

\onlineid{1245}

\vgtccategory{Research}
\vgtcpapertype{Application/Design study}

\title{Visual Analysis of Discrimination in Machine Learning}
\author{Qianwen Wang, Zhenhua Xu, Zhutian Chen, Yong Wang, Shixia Liu, and Huamin Qu}
\authorfooter{
\item
 Qianwen Wang, Zhenhua Xu, Zhutian Chen, Yong Wang, and Huamin Qu are with Hong Kong University of Science and Technology.\\ E-mail: \{qwangbb, zxubg, zhutian.chen, ywangct\}@connect.ust.hk, huamin@ust.hk.
\item
 Shixia Liu is with Tsinghua University. E-mail: shixia@tsinghua.edu.cn.
}

\shortauthortitle{Biv \MakeLowercase{\textit{et al.}}: Global Illumination for Fun and Profit}
\abstract{%
The growing use of automated decision-making in critical applications, such as crime prediction and college admission, has raised questions about fairness in machine learning.
How can we decide whether different treatments are reasonable or discriminatory?
In this paper, we investigate discrimination in machine learning from a visual analytics perspective 
and propose an interactive visualization tool, DiscriLens, to support a more comprehensive analysis.
To reveal detailed information on algorithmic discrimination, 
DiscriLens identifies a collection of potentially discriminatory itemsets based on causal modeling and classification rules mining.
By combining an extended Euler diagram with a matrix-based visualization, we develop a novel set visualization to facilitate the exploration and interpretation of discriminatory itemsets.
A user study shows that users can interpret the visually encoded information in DiscriLens quickly and accurately.
Use cases demonstrate that DiscriLens provides informative guidance in understanding and reducing algorithmic discrimination.
} % end of abstract

\keywords{Machine Learning, Discrimination, Data Visualization.}

\CCScatlist{ % not used in journal version
 \CCScat{K.6.1}{Management of Computing and Information Systems}%
{Project and People Management}{Life Cycle};
 \CCScat{K.7.m}{The Computing Profession}{Miscellaneous}{Ethics}
}

\teaser{
  \centering
  \includegraphics[width=\linewidth]{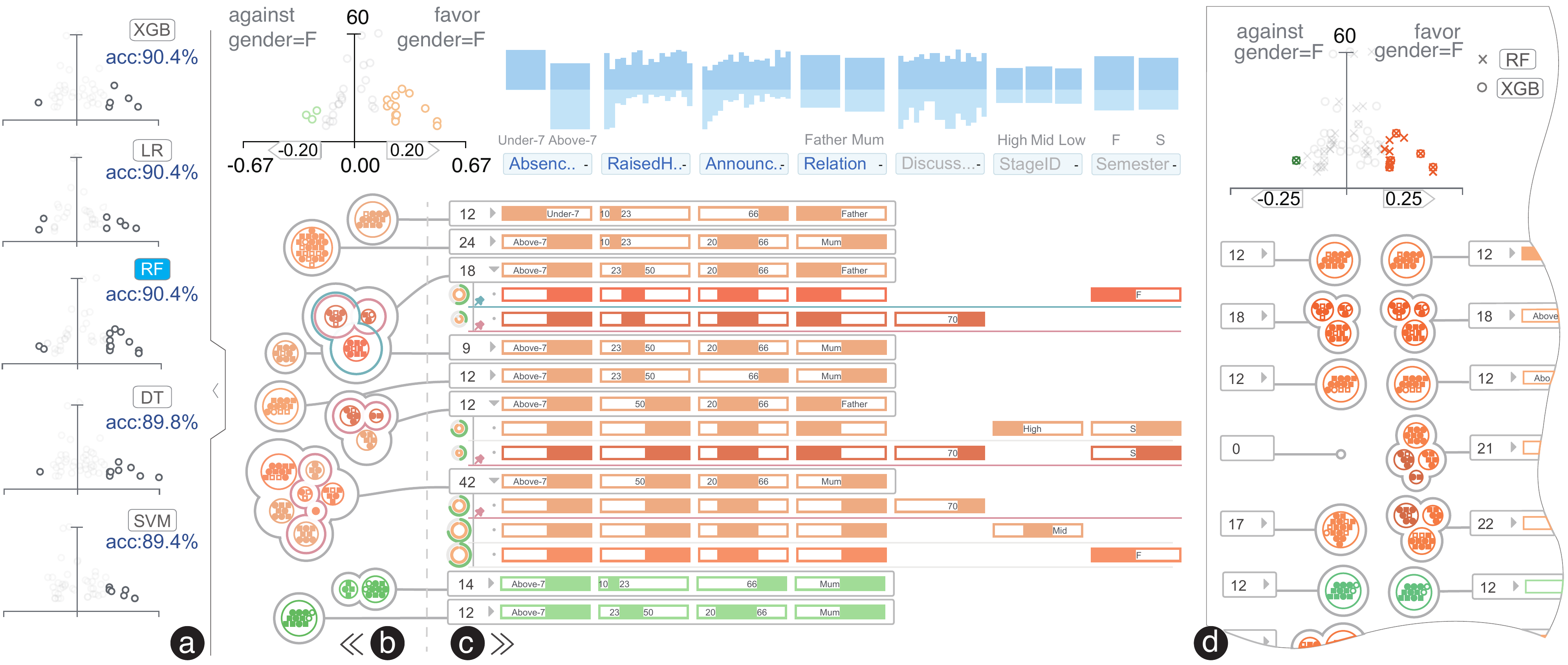}
  \vspace{-5mm}
  \caption{DiscriLens facilitates a better understanding and analysis of algorithmic discrimination: (a) scatter plots offer an overview of the discriminatory itemsets; (b) RippleSets reveal the intersections among these itemsets; (c) the attribute matrix represents the details of each discriminatory itemset; (d) the comparison mode enables users to compare two models side by side. }
	\label{fig:teaser}
}

\vgtcinsertpkg

\begin{document}


% \firstsection{Introduction}

\maketitle

\input{sections/1.introduction.tex}
\input{sections/2.related_work.tex}

\input{sections/3.background.tex}

\input{sections/4.system.tex}
\input{sections/5.discrimination_discovery.tex}

\input{sections/6.visual_design.tex}

\input{sections/7.evaluation.tex}

\input{sections/8.discussion_conclusion.tex}

\acknowledgments{
This research was supported in part by Hong Kong Theme-based Research Scheme grant T41-709/17N and also a grant from MSRA.
}

\bibliographystyle{abbrv-doi}

\bibliography{ref}
\end{document}

%% file: sections/1.introduction.tex
\section{Introduction}
Machine learning (ML) has progressed dramatically in recent decades and become a useful technique in a variety of applications, 
including credit scoring~\cite{khandani2010consumer}, crime prediction~\cite{gerber2014predictingCrime}, and college admission~\cite{ragab2014comparative}.
Since decision-making in these areas may have ethical or legal issues~\cite{eu-equal, uk-equal}, it is crucial for model users to go beyond model accuracy and consider the fairness of ML models.
% In this study, we study discrimination in binary classification problems.

Consider the following scenario.
When reviewing loan applications, loan officers need to estimate the risk of default (\ie, the probability of failing to repay the loans),
which is usually time-consuming and error-prone. 
A machine learning model trained on historical credit data can estimate the creditworthiness of applicants and thus facilitate the decision-making.
However, this model can unintentionally make discriminatory predictions in the social sense, even though the training data describes objective facts and includes no human discrimination.
For example, this model may treat two applicants unequally based on gender despite their same repayment capacity.
% These discriminatory predictions can hardly be eliminated
To avoid making decisions based on protected attributes (attributes such as gender and race that are legally protected by laws~\cite{uk-equal, us-equal}),
a straightforward method is to hide these attributes. 
But this method not only decreases the model accuracy but has also been proven ineffective since models are able to learn protected attributes from other non-protected attributes (\eg, predict gender based on address and occupation)~\cite{Hajian2016AB, pedreshi2008discrimination,  vzliobaite2011handling}.

% Consider the following scenario.
% When reviewing college applications, officers need to estimate the academic capacity of the students based on a series of attributes (\eg, SAT). 
% A machine learning model trained on historical data can estimate the capacity of applicants and thus facilitate the decision-making~\cite{kusner2017counterfactual}.
% However, this model can unintentionally make discriminatory predictions in the social sense.
% For example, this model may treat applicants unequally based on race despite their same academic capacity.
% % These discriminatory predictions can hardly be eliminated
% To avoid decisions based on protected attributes (attributes that are legally protected by laws~\cite{uk-equal, us-equal}),
% a straightforward method is to hide these attributes. 
% However, this method not only decreases the model accuracy but also has been proven ineffective, since models are able to predict protected attributes from other non-protected attributes (\eg, predict race based on address)~\cite{pedreshi2008discrimination, Hajian2016AB, vzliobaite2011handling}.

To further promote the adoption of ML models and prevent potential negative social impacts, discrimination in ML is drawing increasing research attention.
% % Previous studies~\cite{Hajian2016AB} suggest two main resources of algorithmic discrimination: the training data and the training algorithm.
% On the one hand, human discrimination may exist in the historical records ~\cite{mancuhan2014combating,pedreschi2009measuring,kamiran2012data}.
% The model may treat human discrimination as general rules and learns to mimic such practices.
% % Various methods have been proposed to remove human discrimination in the training data~\cite{mancuhan2014combating,pedreschi2009measuring,kamiran2012data}.
% On the other hand, even if the training data is free from human discrimination, discriminatory predictions can be introduced by the learning mechanism (\eg, over-fitting)~\cite{zhang2017achieving, Hajian2016AB}.
% % Algorithmic discrimination, either caused by the training data or the training process, 
% % requires a comprehensive analysis and a detailed understanding to better guide evaluation and improvement.
Many methods have been proposed to assess and mitigate discrimination from three main aspects: pre-process methods that investigate the discrimination in training data~\cite{kamiran2012data, kusner2017counterfactual, zhang2017achievingData}, 
in-process methods that adjust the model learning process~\cite{kamishima2012fairness, mancuhan2014combating, zafar2015fairness},
and post-process methods that modify the discriminatory model predictions~\cite{hardt2016equality, zhang2017achieving}.
However, these studies usually formalize discrimination as summary statistics and may hinder a detailed assessment.
Meanwhile, these studies simply assume that the representation of discrimination has been clearly defined, which usually does not hold in practice ~\cite{pedreschi2012study, hardt2016equality}.
Due to the complex nature of discrimination, it has no clear and uniform definition and its representation varies a lot in different domains.
In this study, we develop a visual analysis tool that enables the involvement of domain knowledge and supports a systematical assessment of discrimination, thus further benefiting the analysis and mitigation of discrimination~\cite{holstein2019improving}.

\begin{figure}[t]
    \centering
    \includegraphics[width=\linewidth]{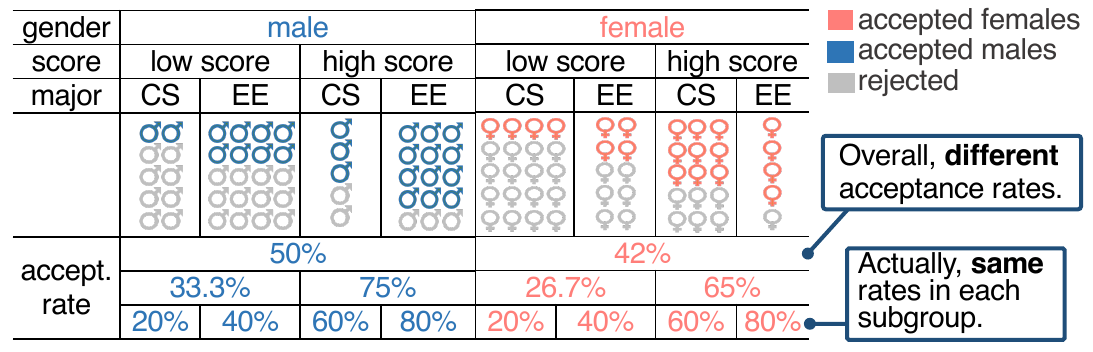}
    \vspace{-2em}
    \caption{It is not easy to distinguish reasonable differences from discriminatory treatment. Here is a toy example of college admission. The overall low acceptance rate for females can be explainable by their tendency to apply to the more competitive major.}
    \label{fig:example}
    \vspace{-1em}
\end{figure}

% We focus on discrimination in binary classification problems. 
To analyze discrimination in machine learning, it is important to access whether differential treatment is discriminatory (\eg, based on race) or reasonable (\eg, based on the qualification of the applicant).  
The term \textit{fairness} used in this paper refers to the principle that any two individuals who are \textit{similar with respect to a particular task} should be \textit{classified similarly}~\cite{Dwork2012fairness, Kilbertus2017avoiding}.
The first question is \textbf{which individuals should be regarded as similar for the task at hand}.
The definition of similar people varies a lot among tasks and is important for the analysis of discrimination. \autoref{fig:example} shows an example of college admission,
% \footnote{This example is also used in \cite{bickel1975sex, zhang2017achievingData, vzliobaite2011handling}} 
where \texttt{gender} is a protected attribute.
If all the applicants are treated as similar, there seems to be gender discrimination since the acceptance rate is 42\% for females but 50\% for males.
Unequal treatment still exists if we group applicants based on the \texttt{test score}.
But if we define similar people using the combination of \{\texttt{major}, \texttt{test score}\}, it then shows no unequal treatment in any of the sub-groups.
Such a phenomenon is also mentioned in Simpson’s paradox~\cite{SimpsonParadox}.
% In other words, the overall low admission rate for females is not discrimination since it can be explainable by females' tendency to apply to more competitive majors~\cite{bickel1975sex, vzliobaite2011handling}.
The second question is \textbf{how to present discrimination among these similar individuals effectively}.
We treat a group of similar people as an \textbf{itemset} defined by a series of attributes values (\eg, \{\texttt{test score=low, major=CS}\}). 
The interpretability of these itemsets is severely weakened when the definition is long and complex.
Furthermore, the number of these itemsets can be large and these itemsets are often intricately intertwined together.
Therefore, it is non-trivial to assist users in perceiving these itemsets and interpreting discrimination. \looseness=-1

\hypersetup{urlcolor=black} % change url color here
To tackle the challenges, we design and implement DiscriLens, an interactive visualization tool that facilitates an easy interpretation, evaluation, and comparison of the algorithmic discrimination.
A demo is available at \url{https://qianwen.info/demos/discrilens/} (recommend opening in Chrome).
We develop a three-stage pipeline to identify a collection of potentially discriminatory itemsets based on causal modeling and classification rules mining.
A set of user interactions are provided to incorporate human domain knowledge in discrimination analysis.
A novel Euler-based visualization, RippleSet (\autoref{fig:teaser}(b)), is proposed to provide an effective presentation of discrimination. 
RippleSet represents one set as several adjacent circles instead of one convex shape, thereby preventing 
% the complex overlapping zones 
the overlaps in traditional Euler diagram.
We further combine the RippleSet with a matrix-based visualization (\autoref{fig:teaser}(c)) to support users in examining the discriminatory itemsets from multiple aspects.
% In addition, a set of user interactions is developed to enable a smooth transition from the overview to the details.
We demonstrate the effectiveness of DiscriLens in analyzing discrimination through a user study and use cases.

The main contributions of this work are as follows:
\begin{itemize}
\setlength\itemsep{0.1em}
    % \item A novel set visualization approach that combines an extended Euler diagram with a matrix-based set visualization.
    % \item A characterization of discrimination in machine learning from a visual analysis perspective, and a thorough discussion of the design goals.
    \item The design and development of DiscriLens, an interactive visual analysis tool with a set of novel visualization techniques for analyzing discrimination in machine learning.
    \item A user study and a series of use cases that assess the utility and usability of DiscriLens. 
    % \item Case studies and lessons learned during the design, implementation, and evaluation of DiscriLens.
\end{itemize}

% \setlength\tabcolsep{3.5pt}
% \begin{table}[]
% \small
% \caption{The overall low admission rate for females can be explainable by their tendency to apply to more competitive majors.}
%     \label{tab:toy_example}
%     \centering
%     \begin{tabular}{*{8}{c|}c}
%     \toprule
%          Gender & \multicolumn{4}{c|}{Female } & \multicolumn{4}{c}{Male } \\
%          \hline
%       Test Score & \multicolumn{2}{c|}{ Low } & \multicolumn{2}{c|}{ High  } & \multicolumn{2}{c|}{ Low } & \multicolumn{2}{c}{ High }  \Tstrut\Bstrut\\
%          \hline
%          Major & CS & ECE  & CS & ECE  & CS & ECE  & CS & ECE \Tstrut\\
%          Applicants & 450 & 150 & 300 & 100 & 150 & 450  & 100 & 300 \\
%          \hline
%          \rowcolor{lightOrange}
%          \multirow{ 3}{*}{Rate} & 20\% & 40\% & 50\% & 70\% & 20\% & 40\% & 50\% & 70\% \Tstrut\\
%          \cline{2-9}
%           & \multicolumn{2}{c|}{25\%} & \multicolumn{2}{c|}{35\%} & \multicolumn{2}{c|}{55\%} & \multicolumn{2}{c}{65\%} \Tstrut\\
%          \cline{2-9}
%          & \multicolumn{4}{c|}{ 29\% } & \multicolumn{4}{c}{59\%  } \Tstrut\\
%     \bottomrule
%     \end{tabular}
%     \vspace{-6mm}
% \end{table}

%% file: sections/2.related_work.tex
\section{Related work}
% The related work of this paper can be categorized into three types: 
% discrimination in machine learning, model-agnostic visual analysis of model behavior, and set visualization.

\subsection{Discrimination in Machine Learning}
% With the adoption of machine learning models in real world applications (\eg, loan approval), machine learning discrimination has gained increasing attention over the recent years.
% Machine learning discrimination mainly comes from two sources: the training data and the trained model. 
Existing studies mainly investigate algorithmic discrimination from two aspects: the data and the model training process~\cite{Hajian2016AB}. 

% On the one hand, 
Training data may include human discrimination and then influence the trained model~\cite{kusner2017counterfactual}.
% trained on the data. 
Various approaches have been proposed to discover discrimination existing in the training data~\cite{chen2020,kamiran2012data, mancuhan2014combating,pedreschi2009measuring,vzliobaite2011handling}. 
% A straightforward solution is to examine different treatments among similar
A pragmatic solution is to examine different treatments among similar
items~\cite{kamiran2012data, pedreschi2009measuring}.
For example, Luong et al.~\cite{luong2011k} used the k-nearest neighbor algorithm to group similar items and identify the discrimination among them.
Pedreshi et al.~\cite{pedreshi2008discrimination} employed classification rules to reveal itemsets that may lead to unequal treatment.
Recently, \qianwen{to provide an interpretable notation of discrimination}, 
several studies considered the relationships among item attributes and examined discrimination from a causal modeling perspective~\cite{kusner2017counterfactual, zhang2017achievingData}. \looseness=-1

Even though human discrimination can be identified and eliminated from the training data, 
a model can still make discriminatory predictions due to the data distribution and the model learning mechanism  (\eg, over-fitting)~\cite{liu2017towards,pedreschi2012study, yuan2021survey,zhang2017achieving, vzliobaite2011handling}.
To remove discrimination introduced in the training process, researchers developed regularizers that penalize discriminatory predictions~\cite{kamishima2012fairness, zafar2015fairness} and methods that directly modify the model predictions~\cite{hardt2016equality, zhang2017achieving}.
Many studies have defined different criteria for model discrimination to provide meaningful~\cite{pedreschi2012study, zhang2017achieving} and interpretable~\cite{hardt2016equality} notations.
Meanwhile, to support users in measuring and comparing discrimination in different models,
initial efforts have been made to quantify model discrimination~\cite{speicher2018unified}. 

% The discrimination analysis in this paper is based on the aforementioned studies. 
% % Moreover, 
% We utilize visualization techniques to facilitate an easy interpretation and a comprehensive analysis of the discrimination, which is not supported by previous studies.
\qianwen{
These studies shed light on discrimination analysis and form the foundation of this study.
However, the discrimination mined by these methods can be complex and thus hard to interpret (\eg, a large number of rule lists), requiring the support of effective visual presentation.
More importantly, the analysis of discrimination is, by nature, a subjective task and varies based on the application domains. 
This paper provides interactive analysis to help incorporate human domain knowledge in discrimination analysis.
}

% To handle this issue, in our paper, a visualization system is designed to assist users to explore more details of discriminant discovery to better understand and compare model discrimination.

\subsection{Visual Analysis for ML Discrimination}
\label{subsec:discrimination_vis}

Visualization, an effective tool for information communication, has been employed to present and analyze discrimination in ML~\cite{ahn2019fairsight, attacking-discrimination-in-ml, cabrera2019fairvis, aif360, fairnessID, fairlearn}.
For example, Google Big Picture developed an interactive visualization demo~\cite{attacking-discrimination-in-ml} to illustrate the discrimination in ML decisions. 
\qianwen{
Other tools, such as IBM AI Fairness 360~\cite{aif360}, Tensorflow Fairness Indicators~\cite{fairnessID}, and FairLearn, enable the easy computation of fairness metrics and support visualizations of these fairness metrics.
However, most tools only support segmenting groups based on one protective attribute and only provide basic visualizations (\eg, bar charts).
Therefore, they cannot be directly applied to investigate and present the potentially complicated discrimination between different groups.
}

Recently, advanced visualization tools have been developed to enables more comprehensive analysis of discrimination, including FairSight~\cite{ahn2019fairsight} and FairVis~\cite{cabrera2019fairvis}. 
FairSight represents a workflow that supports the four fairness-aware actions (\ie, understand, measure, identify, and mitigate) required in decision making.
FairSight shares the same high-level goal as DiscriLens and also supports the analysis of individual fairness.
However, FairSight either a) identifies discriminated instances based on an KNN algorithm or b) provides a global-level measure by aggregating over all instances. As a result, FariSight fails to uncover the discriminatory itemsets and to reveal \textit{``when and where will a model yield discriminatory predictions''}.
FairVis is more related to DiscriLens due to its focus on the analysis of discriminatory itemsets.
This tool helps users to audit the fairness among different itemsets.
However, FairVis focuses on suggesting similar discriminatory itemsets during analysis. 
The possible complex definition of the itemsets are presented using tables, which is not an optimal representation method. No special visualizations are designed to reveal the intertwining relationships among the subgroups. 

In DiscriLens, we propose novel visualizations for better understanding and analysing ML discrimination.

%--------------------------------------------------------------------
\subsection{Set Visualization}
The core of discrimination analysis is to understand how a set of items, grouped by attribute values, are treated unequally. 
Due to the ubiquity of set-based analysis, a large body of literature has been developed.
Below we review some visualizations that are germane to our work.
A comprehensive review of set visualizations can be found in the work of Alsallakh et al.~\cite{alsallakh2014setsurvey}. 

The most common set visualizations are Euler and Venn diagrams~\cite{freiler2008interactive, lex2014upset}, 
which represent sets as convex or concave shapes and show set intersections as overlapping shapes.
% which use shape boundaries to indicate set memberships.
% \emph{Euler} diagrams represent each set as a geometric shape and show the intersections by overlapping these shapes. 
% \emph{Venn} diagrams can be treated as a special form of Euler diagrams that show all intersections. 
% However, complex set memberships can result to diagrams that are difficult to interpret.
% Various optimization methods have been proposed~\cite{riche2010untangling}.
Previous work has proposed many extensions of Euler diagram with varying design goals~\cite{collins2009bubbleset, Dinkla2012kelp,riche2010untangling, simonetto2009fully}.
For example, Riche et al.~\cite{riche2010untangling} simplified a sophisticated collection of intersecting sets into a strict hierarchy, thereby untangling Euler diagrams and improving its readability.
% "consider the set relation as primary and create a spatial layout according to set membership"
% use continuous isocontours to 
Bubble Sets~\cite{collins2009bubbleset} delineate set memberships and minimize cluster overlap while retaining the layout that reflects the semantic spatial organization.
However, methods based on Euler and Venn diagrams often impose severe limitations on the number of sets and the complexity of set intersections~\cite{Alsallakh2013radicalset}.
% In this study, we extend the Euler diagram and propose a novel set visualization, RippleSet, to help users better interpret complex set relationships.
% The resulting diagrams can be difficult to interpret when the set relationships are complex.

% Apart from extending Euler and Venn diagrams, researchers have developed many alternative presentations of set data~\cite{alsallakh2017powerset, sadana2014onset, lex2014upset, Alsallakh2013radicalset}.
% Linesets \cite{alper2011linesets} represents a set as a curve that connects all items of this set, improving the
% readability of set memberships and intersections.
% Alsallakh et al~\cite{alsallakh2017powerset} developed Powerset, a visualization technique based on tree maps, to provides a comprehensive overview of all set intersections.
% Parallel Sets ~\cite{kosara2006parallelset} adopt the flexible layout of parallel coordinates and visualize categorical sets as the parallelograms between axes.
% % Onset~\cite{sadana2014onset} for large-scale binary set data

Considering the inherent limits of Euler and Venn diagrams, researchers have developed many alternative presentations of set data, including curve-based presentation~\cite{alper2011linesets}, treemap-based presentation~\cite{alsallakh2017powerset}, parallel coordinate-based presentation~\cite{kosara2006parallelset}, matrix-based presentation~\cite{Kim2007conset, lex2014upset}, and chord diagram-based presentation~\cite{Alsallakh2013radicalset}. The most relevant ones to our work are Upset~\cite{alsallakh2017powerset} and Conset~\cite{Kim2007conset}, which employed matrix to communicate the properties of the set aggregations and intersections. 
However, Upset and Conset are specified for visualizing binary set data and cannot be directly applied to the categorical set data in discrimination analysis.
Meanwhile, in existing work, the comparison of two collections of set data has not been adequately explored and requires further investigation.

In this work, to facilitate the analysis and comparison of categorical set data, we propose a novel set visualization that combines an extended Euler diagram, RippleSet, with a matrix-based visualization.\looseness=-1

% Upset~\cite{lex2014upset} effectively communicates the properties of the set aggregations and intersections.

% Our work is built upon the previous work on set visualization.

%% file: sections/3.background.tex
\vspace{-1mm}
\section{Discrimination: A Mathematical Notation}
% \section{Problem Formulation}
This section provides background information and a mathematical notation of discrimination.
% In this section, we formalize the mathematical notation of algorithmic discrimination.
% We define a supervised model as $\hat{Y}=f(X, A)$ that predict the ground truth $Y$, where $A$ is the protected attribute (\eg, gender) and $X$ are other attributes.
% Following previous practice~\cite{zhang2017achievingData, kusner2017counterfactual, hardt2016equality}, 
\qianwen{In this paper, we focus on the analysis of a single protected attribute, denoted by $A$.}
We denote other non-protected attributes by $X$, and the outcome by $Y$.
Note that the outcome $Y$ can be the labels in the training data or the predictions of ML models.
The beneficial class (\eg, a loan approval) is denoted by $Y=1$, and the protected group (\eg, females) is denoted by $A=1$.
Unobserved attributes are not considered in this study.

\begin{figure}[t]
    \centering
    \includegraphics[width=0.95\linewidth]{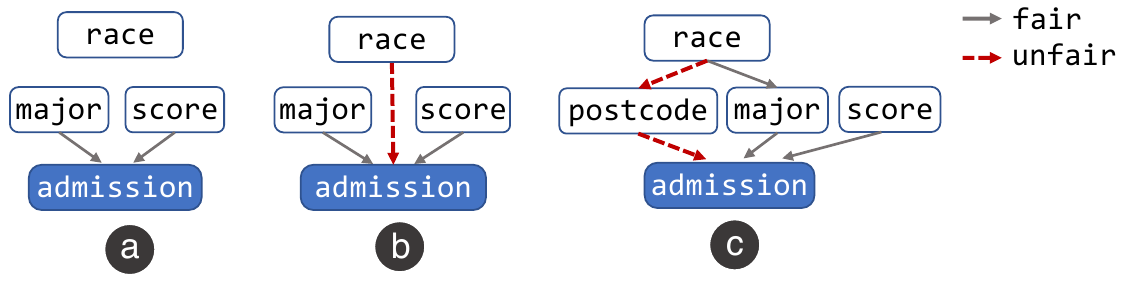}
    \vspace{-1em}
    \caption{In a toy example of college admission, different types of relationship exist between the protected attribute \texttt{race} and the result \texttt{admission}. (a) \texttt{race} doesn't influence \texttt{admission}. (b) \texttt{race} directly influences \texttt{admission}. (c) \texttt{race} indirectly influences \texttt{admission} through other attributes.}
    \vspace{-1em}
    \label{fig:causal_example}
\end{figure}

\qianwen{A natural notation of fairness is demographic parity~\cite{hardt2016equality, kusner2017counterfactual}, which states that each segment of a protected class (\eg, blacks and non-blacks) should receive positive outcomes at equal rates (\eg, same admission rate), regardless of the value distribution of other attributes (\eg, score). 
In other words, this notation treats all individuals as similar and requires the decisions to be independent of the protected attribute, as shown in \autoref{fig:causal_example}(a)(b).
Demographic parity is easy to examine and thus has been widely used in legal practice, including the US Equal Employment Opportunity Commission~\cite{us-employ} and the UK Sex Discrimination Act ~\cite{uk-sex}.
However, this notion can be flawed when the correlation between attributes is strong and needs to be considered, which is the situation we consider in this paper.
}
% However, this notion can be seriously flawed since the protected attribute is usually correlated with other attributes.

As shown in \autoref{fig:causal_example}(c), the protected attribute \texttt{race} are correlated with other attributes and can influence the \texttt{admission} from two paths: \textbf{justified difference} \qianwen{(\texttt{race}$\rightarrow$\texttt{major}$\rightarrow$\texttt{admission})} and \textbf{indirect discrimination} \qianwen{(\texttt{race}$\rightarrow$\texttt{postcode}$\rightarrow$\texttt{admission})}.
This first path is called justified difference since \texttt{major} offers objective information about the \texttt{admission} and can explain the outcome differences. \textit{E.g.}, the overall low admission rate for black students can be explained by their tendency to apply for more competitive majors.
Such attributes that can justify decision differences are called \textbf{resolving} attributes~\cite{Kilbertus2017avoiding}.
The second path is called indirect discrimination because the \texttt{postcode} is correlated with \texttt{race} (\eg, majority-black neighborhood) but should not affect the \texttt{admission}.
These attributes that act as a proxy for protected attributes are called \textbf{proxy} attributes~\cite{Kilbertus2017avoiding}.
As a result, resolving and proxy attributes should be both considered for the analysis of discrimination~\cite{Kilbertus2017avoiding, zhang2017causal}. 
\qianwen{
In general, there is no ground truth for resolving or proxy attributes.
Their definitions depend on the task (\eg, college admission) and require human domain knowledge.
Therefore, interactive visual analysis of discrimination is needed.
}

In this study, we use \textbf{risk difference} (RD), a widely used metric in anti-discrimination literature~\cite{pedreschi2012study, romei2014multidisciplinary, zhang2017achievingData, zhang2017causal}, for the measurement of discrimination. 
$RD = P(Y=1|A=1, S=s_i) - P(Y=1|A=0, S=s_i)$, where $S$ \qianwen{includes} all resolving attributes and no proxy attributes, ensuring that justified difference is separated from discrimination.
Following the practice in~\cite{zhang2017achievingData}, non-discrimination is claimed when $RD<\tau$ holds for each itemset $s_i$. 
Note that DiscriLens is not metric specific and can be easily extended to other metrics such as true positive rate difference, error rate ratio, and odds ratio~\cite{aif360}.
\qianwen{We would like to emphasize that the discrimination measurement used in this paper is just one of the many possible choices. The choice of measurement depends on the context of analysis.}

%% file: sections/4.system.tex
\section{Designing DiscriLens}
% In this section, we first describe the design goals and then give an overview of DiscriLens.

% \subsection{Target Users}
% domain experts with a high level of expertise in the application

\subsection{Design Goals}
The goal of DiscriLens is to enable model users to flexibly interpret and analyze the discrimination in ML. 
The users have a certain level of expertise in ML and wish to apply ML models to a real-world application. 
They want to analyze discrimination to better avoid the potential ethical or legal issues.
% and want to avoid algorithmic discrimination in the application.
Based on the previous literature on algorithmic discrimination and discussions with two ML experts,  
we identify the following design goals.
% to guide the development of DiscriLens.

\begin{enumerate}[label=\textbf{G\arabic*}]
\setlength{\leftmargin}{0pt}

% ----------------------------------
\item \label{goal:validate} \textbf{Customize the definition of discrimination.}

Discrimination is a complex concept and can have different representations at different domains~\cite{naff1995subjective}.
For example, a decision based on age is discrimination in certain domains but not in others.
Also, the resolving attributes to a decision significantly vary from domain to domain.
Therefore, the domain knowledge of the users is essential for the analysis of discrimination.
Users should be allowed to customize the definition of discrimination (\ie, protected attribute, proxy attributes, resolving attributes, the threshold $\tau$) based on their domain knowledge as well as the results of discrimination discovery algorithms.

% ----------------------------------
\item \label{goal:measure} \textbf{Measure the degree of discrimination.}

For the analysis of discrimination, model users need to measure the degree of algorithmic discrimination~\cite{pedreschi2012study, speicher2018unified}, which can be evaluated by two indicators: the \textit{risk difference} and \textit{size}.
The \textit{risk difference} answers the question: \textit{``To which extent are the protected group \qianwen{($A=1$)} and non-protected group  ($A=0$) treated unequally?''}
The \textit{size} answers the question: \textit{``How many people are influenced by algorithmic discrimination?''} High \textit{risk difference} and large \textit{size} indicate severe discrimination.

% ----------------------------------
\item \label{goal:condition} \textbf{Identify the condition of discrimination.}

Apart from measuring the degree of discrimination, the condition of discrimination should be identified to reveal when a model will yield discriminatory predictions.
% This information can be obtained by identifying the discriminatory itemset (\ie, a series of attribute values).
This information can be conveyed by a set of attribute values that specify the discriminatory itemsets, \ie, a group of individuals who are similar to the task but are treated differently.
The condition of discrimination is crucial since it can provide detailed guidance on applying models (\eg, identify the failure mode~\cite{pedreshi2008discrimination}), improving the training data
(\eg, modify inappropriate data labels~\cite{zhang2017achievingData}, collect new data~\cite{Zhang2018examining}), and adjusting the training process (\eg, add regularization to penalize discriminatory predictions~\cite{zafar2015fairness, mancuhan2014combating}).

% ----------------------------------
\item \label{goal:distribution} \textbf{Depict the distribution of discrimination. }
\qianwen{
The degree and the condition of discrimination can vary a lot among itemsets~\cite{pedreshi2008discrimination,  vzliobaite2011handling}.
For example, in income prediction, the degree of discrimination towards blacks may vary with other attribute values, such as \texttt{working hours}, \texttt{education level}, and \texttt{work class}. 
As a result, algorithmic discrimination can hardly be comprehensively described by summary statistics or easily interpreted by humans~\cite{speicher2018unified}.
% The difficulty is significantly increased when the relationships between items (\eg, intersection, inclusion) are complex.
The difficulty is significantly increased when the intersections between itemsets are complex.
To fully capture the discriminatory behavior of a machine learning model, the distribution of discrimination should be offered to help users understand how discrimination varies among and inside different itemsets. \looseness=-1
}
% ----------------------------------
\item \label{goal:compare}
\textbf{Compare discrimination. }
\qianwen{
Summary statistics can be ineffective for the analysis of ML models~\cite{yuan2021survey}.
Models with the same accuracy can assign conflicting predictions for the same input data~\cite{marx2019predictive}, leading to discriminatory predictions against different groups of data items.
For example, one model may discriminate equally against all female students, while another model discriminates severely against females with low scores but mildly against other females.
Therefore, to choose an appropriate model, it is important to compare models and analyze conflicting predictions.}

\qianwen{
Apart from model-level comparison, it is also important to compare discrimination among different itemsets. 
Several studies~\cite{pedreschi2012study, corbett2017algorithmic} have observed that model discrimination cannot be completely removed. 
Therefore, different itemsets should be compared to prioritize the most severe discrimination, whose identification is objective and requires human domain knowledge.
}
% The measurement of severity requires human domain knowledge since it is a combination of multiple variables such as risk differences, size, and condition.
% A comparison among itemsets should be supported to help users identify the relatively more severe discrimination.

% First, users need to compare discrimination among different itemsets. When taking measures to remove discrimination, discriminatory predictions are usually penalized/modified differently~\cite{kamiran2012data, kamiran2012decision}. 
% Therefore, priority should be given to the most severe discrimination.
% The measurement of severity requires human domain knowledge since it is a combination of multiple variables such as risk differences, size, and condition.
% % A comparison among itemsets should be supported to help users identify the relatively more severe discrimination.

% Second, users need to compare discrimination among different machine learning models.
% Several studies~\cite{pedreschi2012study, corbett2017algorithmic} have observed that it is infeasible for a model to perfectly satisfy fairness conditions (just as it is impractical to obtain a 100\% accurate model).
% Hence in practice, model users usually need to compare discriminatory models and select one from them.

\end{enumerate}

\vspace{-2em}
\subsection{System Overview}

\begin{figure}[t]
    \centering
    \includegraphics[width=\linewidth]{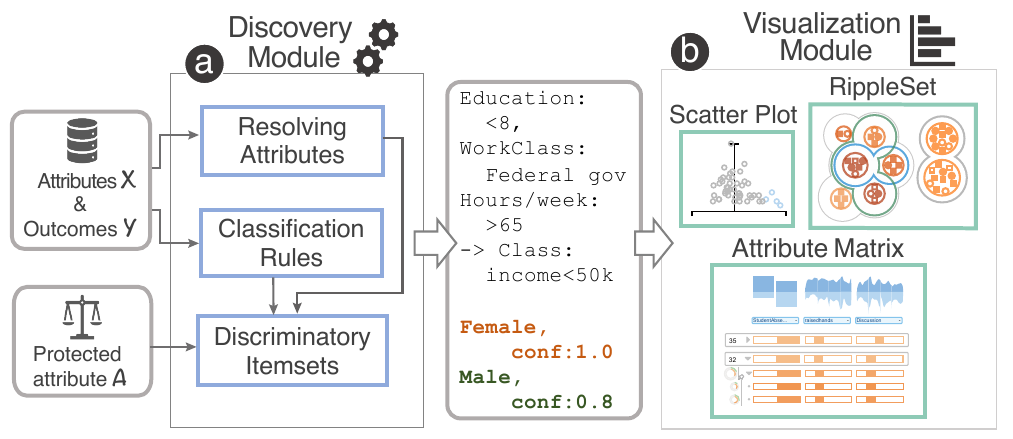}
    \vspace{-2em}
    \caption{DiscriLens consists of two main modules: a discovery module (a) and a visualization module (b).}
    \label{fig:overview}
    \vspace{-4mm}
\end{figure}

DiscriLens consists of two main modules: a discovery module and a visualization module (\autoref{fig:overview}).
\textbf{The discovery module} takes the training data, the model, and the user-defined protected group as input. It then goes through a three-stage pipeline and produces a collection of potentially discriminatory itemsets.
\textbf{The visualization module} serves as an interface that helps understand the discrimination, as well as a tool that provides guidance on applying and improving the model.
% The details of the discovery module are discussed in \autoref{sec:discovering} and the details of the visualization module are discussed in \autoref{sec:visual_interface}.
Details of the discovery and the visualization module are introduced in the following two sections.

%% file: sections/5.discrimination_discovery.tex
\begin{figure}[b]
    \centering
    \vspace{-1em}
\includegraphics[width=0.95\linewidth]{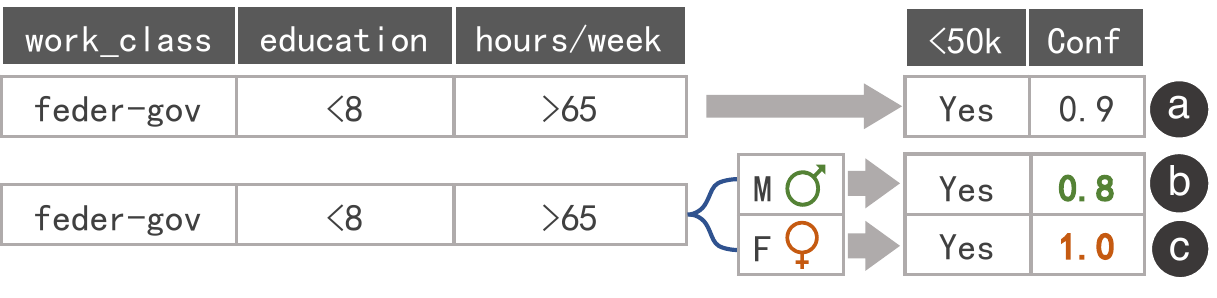}
\vspace{-1em}
\caption{(a) is an example of classification rules. The confidence difference between rule (b) and (c) might indicate a discrimination towards females and requires further examination.}
    \label{fig:rule}
\end{figure}

\section{Discrimination Discovery}
\label{sec:discovering}

We view the task of discovering discrimination as a problem of mining discriminatory itemsets~\cite{pedreshi2008discrimination, zhang2017achievingData}.
Given the outcomes $\hat{Y}=f(X)$ and a threshold $\tau$, the goal is a collection of itemsets $\{ s_1, s_2, ..., s_n \}$ that $ | P(\hat{Y}=1|s_i, A=0)-P(\hat{Y}=1|s_i, A=1) | > \tau, i \in n $, where the itemset $s_i$ is a group of items that are similar to the task.
As shown in \autoref{fig:overview}(a), the discovery module contains three main components:
mine resolving attributes, mine classification rules, and mine discriminatory itemsets.

% The discovering algorithm contains three main steps (\autoref{fig:overview}(a)).
% First, we mine candidate key attributes to guide the following discrimination discovery.
% Second, based on the model predictions, we train a list of classification rules to depict the given model.
% Finally, we identify a collection of potentially discriminatory itemsets in accordance with the classification rules and the key attributes.
% Users are allowed to modify the key attributes.

% The discovering algorithm is based on ~\cite{zhang2017achievingData}.
% Meanwhile, we also adopt the rule-list-based discrimination discovery method proposed in ~\cite{pedreshi2008discrimination} to provide more information about the discrimination distribution among itemsets.

% \subsection{Causal Modeling}
\subsection{Resolving Attributes}
In this step, we mine resolving attributes following the causality-based method proposed in~\cite{zhang2017achievingData}.
Causal modeling is a complicated task and requires strong prior knowledge.
Fortunately, when analyzing discrimination, 
we only need to mine local causality and identify the parents of $Y$ (\ie, $(Pa(Y)$) without building the complete causal graph. 
We refer the readers to \cite{zhang2017achievingData, Kilbertus2017avoiding} for more details and the complete mathematical proof.
\qianwen{
The resolving attributes can be then represented as $Par(Y)\setminus (A \cup P)$, where $P$ indicates the proxy attributes. 
The identification of proxy and resolving attributes requires users' domain knowledge. Therefore, we allow users to interactively modify resolving attributes, remove proxy attributes, and analyze the results in the visualization module.
In this implementation, we use the Fast Greedy Equivalence Search~\cite{chickering2002bayesSearch} for local causal discovery.
}

\subsection{Classification Rules}
% We refer the reader to ~\cite{yin2003cpar} for a
% discussion of the integration of classification and association
% rule mining.
% Given the synthetic samples and the model predictions, 
In this stage,
we can extract a list of classification rules to explain the outcome $Y$.
Take a support vector machine trained on Adult dataset~\cite{uci2019} as an example.
The task is to predict whether the annual income of a person is over 50k USD.
One of the classification rules is shown in \autoref{fig:rule}(a).
The support of this rule $supp(s \rightarrow \hat{Y}=1) = supp(s, \hat{Y}=1) = 16$
and the confidence of the rule $conf(s \rightarrow \hat{Y}=1)=P(\hat{Y}=1|s)=supp(s, \hat{Y}=1)/supp(s)=0.9$.
In this implementation, we use minimum description length~\cite{fayyad1993multi} to discretize continuous attributes and FP-Growth~\cite{han2000fp-growth} to mine the classification rules.
Alternative methods can also be used, and we refer the reader to~\cite{yin2003cpar} for a detailed discussion. \looseness=-1

\subsection{Discriminatory Itemsets}
Given two classification rules: 
$(s,A=0) \rightarrow \hat{Y}=1,
(s, A=1) \rightarrow \hat{Y}=1$, 
we can treat $s$ as a discriminatory itemset if $|P(\hat{Y}=1|A=0, s) - P(\hat{Y}=1|A=1, s) |> \tau$ and $s$ includes all resolving attributes and no proxy attributes.
We do not specify the sign due to reverse discrimination, \ie, discrimination against the non-protected group.

As shown in \autoref{fig:rule}, if \texttt{[workclass, education, hours per week]} include all resolving attributes and no proxy attributes, the confidence difference between rule (b) and (c) can be treated as discrimination towards females.
In other words, in the itemset $s=$\texttt{[workclass=feder-gov, education<8, hours per week>65]},
males are less likely to be predicted to have low incomes compared with their female counterparts (0.8 vs. 1.0).
Therefore, this itemset is called discriminatory itemset.
To enable interactive modification, we do not specify resolving and proxy attributes when mining classification rules. 
Note that $s$ may contain attributes which are neither proxy nor resolving attributes. 
We call these attributes as \textbf{context} attributes. 
In real-world, a discriminatory itemset can be defined by a long list of attribute values, and the itemsets are usually intertwined together.
As a result, effective visualization is needed to assist users in the analysis of discrimination.

%% file: sections/6.visual_design.tex
\section{Visual Interface}
\label{sec:visual_interface}
The visualization module consists of three visualization components (\autoref{fig:teaser}).
The \textbf{scatter plot}~(a) offers an overview of the discriminatory itemsets and allows users to filter itemsets.
% The \textbf{RippleSet}~(b) reveals the intersections among itemsets.
% The \textbf{attribute matrix}~(c) represents the details of each itemset.
\qianwen{
\textbf{RippleSet}~(b) and \textbf{attribute matrix}~(c) supplement each other to support a detailed analysis of discriminatory itemsets. 
RippleSet reveals the discrimination among items that belong to intricately intertwined sets, 
while the attribute matrix illustrates the details of each itemset. 
}

\begin{figure*}
    \centering
    \includegraphics[width=0.95\linewidth]{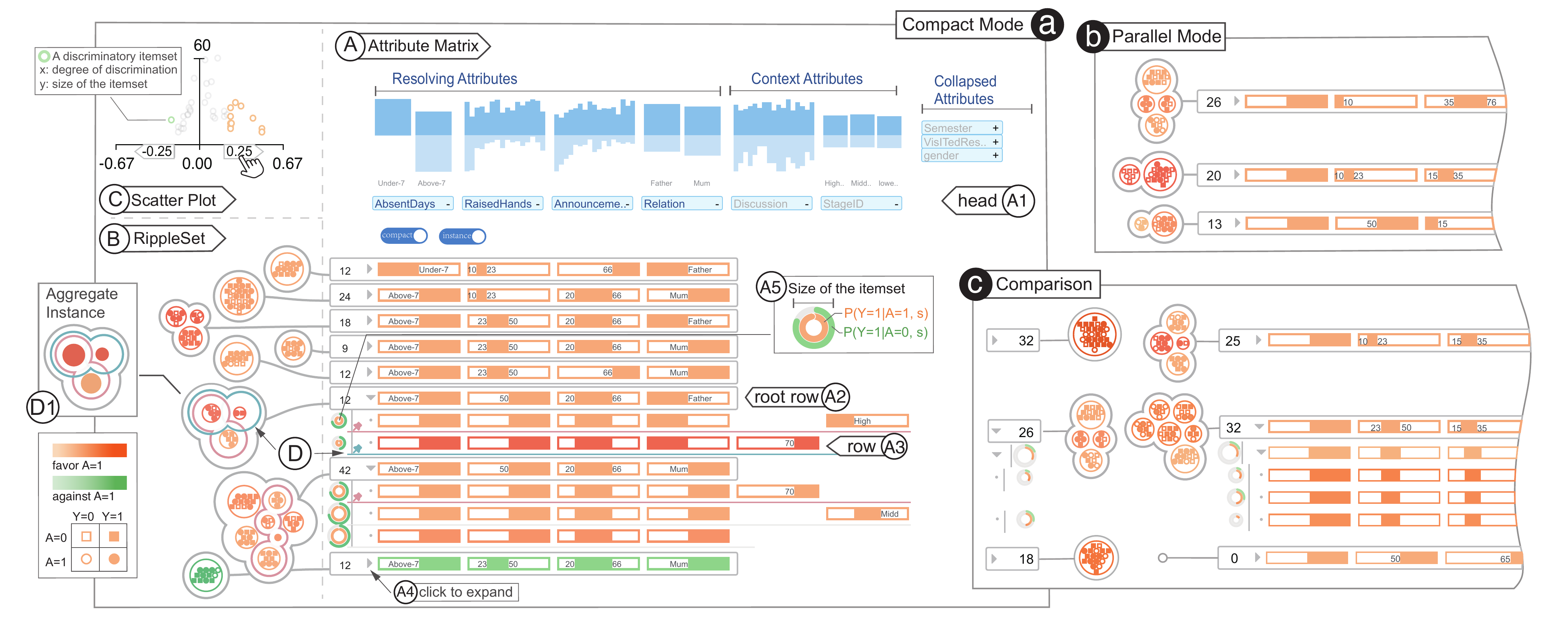}
    \caption{The visual interface consists of three key components: the attribute matrix (A), the RippleSet (B), and the scatter plot(C). Three modes of analysis are supported: the compact mode (a), the parallel mode (b), and the comparison mode (c).}
    \label{fig:interface}
\end{figure*}

\begin{figure}
    \centering
    \includegraphics[width=\linewidth]{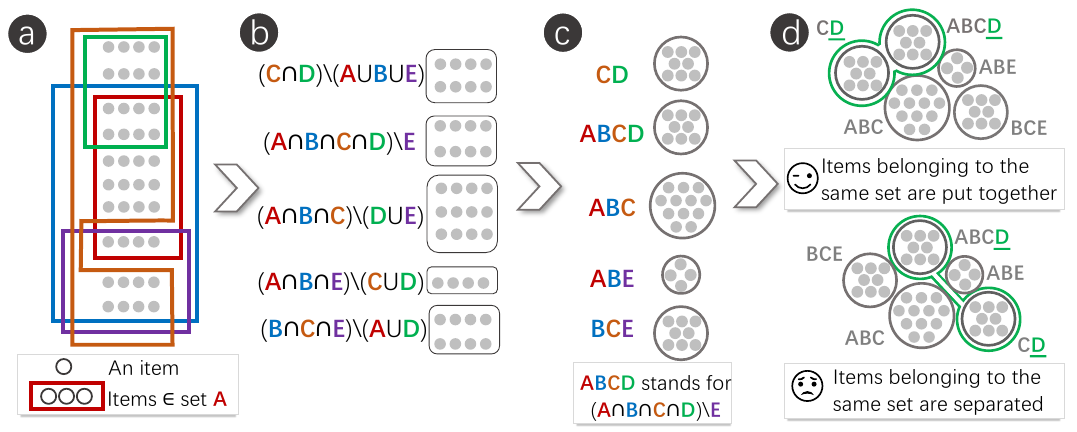}
    \vspace{-2em}
    \caption{RippleSet avoids the complex overlapping shapes in the traditional Euler diagram (a) by representing the maximal inseparable subset (b) as circles (c). Subsets belonging to the same set are put adjacent to one another to better present the composition of one set (d).}
    \label{fig:ripple_formalization}
\end{figure}

\subsection{RippleSet}
\qianwen{
When analyzing model discrimination, it is important to effectively represent discrimination among itemsets that have complex intersection and inclusion relationships (\ref{goal:distribution}).
Existing set visualization methods often impose severe limitations on the complexity of set intersections~\cite{Alsallakh2013radicalset, alsallakh2014setsurvey}, and thus can hardly satisfy the design goals of DiscriLens.  
To better support the analysis in DiscriLens, we extend the widely-used Euler diagram and design RippleSet. \looseness=-1
% To address this issue, we design RippleSet, a set visualization that extends a widely-used set visualization, Euler-diagram. 
}

% % To effectively represent discrimination given the complex intersection and inclusion relationships between itemsets (\ref{goal:distribution}), 
% To effectively represent discrimination among itemsets that have complex intersection and inclusion relationships (\ref{goal:distribution}), 
% we design RippleSet, a set visualization that extends the Euler-diagram. 
% % Each RippleSet represents a collection of discriminatory itemsets and is linked to a root row of the attribute matrix.

\subsubsection{Visual Encoding}

As shown in \autoref{fig:ripple_formalization}, the main idea of RippleSet is to represent one set using several adjacent shapes, instead of one shape, to avoid the complex overlapping shapes in the traditional Euler diagram. 
\qianwen{In RippleSet, each shape represents one maximal inseparable subset, \ie, a subset that will not be further segmented based on the sets they belong to (\autoref{fig:ripple_formalization}(b)).}
Therefore, more shapes in the RippleSet indicate more complex intersections among itemsets.
During the analysis, a set can be highlighted using an outline that surrounds all the subsets of the set.

In RippleSet, we also offer instance-level visualization. 
Each data item is represented as a dot and placed inside the subset it belongs to, as shown in \autoref{fig:interface}(C).
\qianwen{Inside each subset, we use the algorithm proposed by Wang et al.~\cite{wang2006visualization} to layout the data items, which are inserted based on their original order in the dataset. }
\qianwen{
Three visual channels, color, shape, and solid/hollow, are used to display the attributes of data items.
Color, the most effective visual channel, is used to present the most important information, the direction and the degree of the discrimination.
The color encoding is consistent with that used in the attribute matrix: color saturation indicates the absolute value of the risk difference while color hue indicates the discriminated group (protected group or non-protected group).
A solid item represents a data item classified as the beneficial class ($\hat{Y}=1$), whereas a hollow item represents a data item classified as the non-beneficial class ($\hat{Y}=0$).
The shapes of the items encode their groups: circles for the protected group ($A=1$) due to their metaphor of softness and squares for the non-protected group ($A=0$) due to their metaphor of hardness~\cite{liu1997metaphor}.
Even though the shape is a less pop-out visual channel, previous studies \cite{li2009evaluation, wu2010opinionseer, demiralp2014learning} have demonstrated that it can support users in analyzing different item groups. 
Meanwhile, we do admit that the shape encoding may reduce the analysis efficiency. To address this issue, we plan to modify the layout algorithm and double encode the item groups using both the shape and item position in the future.
For a large number of items, RippleSet supports the aggregation of items, as shown in~\autoref{fig:interface}(D1).
More details about the layout algorithm are provided in the supplementary material.}

\begin{figure}
    \centering
    \includegraphics[width=0.9\linewidth]{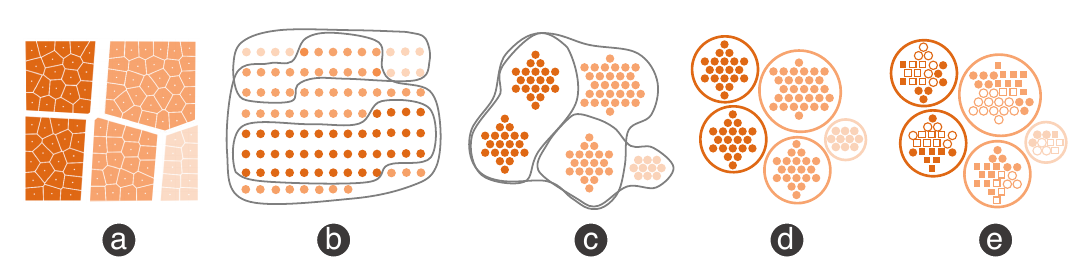}
    \vspace{-4mm}
    \caption{Alternative designs of the set visualization: the weighted Voronoi-based design (a) contains irregular shapes and makes it difficult to compare the size of subsets; the Euler-based design (b) introduces significant visual clutter. We improved the design in (b) through an iterative design process((c)-(e)).}
    \vspace{-1em}
    \label{fig:ripple_alternative}
\end{figure}

\subsubsection{Design Alternatives}
We tested several design alternatives (\autoref{fig:ripple_alternative}) for the set visualization, during which we interviewed domain experts and discussed with visualization experts.
We first tried a weighted Voronoi-based design, as shown in \autoref{fig:ripple_alternative}(a). This design was unsuccessful since the irregular shapes posed challenges to users for comparing the size of subsets.
We then implemented a Euler-based design, as shown in \autoref{fig:ripple_alternative}(b). 
From \autoref{fig:ripple_alternative}(b) to (e), we clustered items belonging to the same subset to offer a clear organization, replaced irregular outlines with circles for aesthetics, and added detailed information of each item.

\subsection{Attribute Matrix}
The attribute matrix (\autoref{fig:interface}(A)) helps users investigate each discriminatory itemset in detail. 
\qianwen{The attribute matrix employs a similar matrix-based layout as RuleMatrix~\cite{ming2018rulematrix}.
This design has been proven effective in presenting classification rules, which is used to identify discriminatory itemsets.}
In the attribute matrix, each row stands for a group of individuals and each column stands for an attribute.
The head (A1) represents the distribution of data items on different attributes, a root row (A2) represents a collection of discriminatory itemsets with the same resolving attribute values, and a row (A3) represents one discriminatory itemset. \looseness=-1

The head (\autoref{fig:interface}(A1)) of the attribute matrix consists of a row of charts.
% that represent the distribution of data items on different attributes.
Histograms are used for visualizing the distribution on continuous attributes, and bar charts are used for discrete attributes.
In each chart, the x-axis represents the attribute value, and the y-axis represents the number of items.
We use dark blue to indicate the items classified as the beneficial class (\eg, a loan approval) and light blue for the non-beneficial class (\eg, a loan decline).
To distinguish resolving attributes, the names of resolving attributes are highlighted.
% Users can open or collapse attributes.
Users can modify the resolving attributes based on their domain knowledge as well as their observation of the data distribution (\ref{goal:validate}).
Discriminatory itemsets will dynamically update according to the modified resolving attributes. \looseness=-1

To facilitate the interpretation and exploration, we group and aggregate discriminatory itemsets to offer a clear summary. 
We first group the discriminatory itemsets based on their resolving attribute values.
All itemsets sharing the same resolving attribute values are put in one collection, which is represented as a bordered rectangle called root row (\autoref{fig:interface}(A2)). The number of items in a collection is labeled at the left end of the bordered rectangle.
Then, in each collection, we organize itemsets, which are represented as rows (\autoref{fig:interface}(A3)), according to their inclusion relations.
% The subset of an itemset, if exists, can be open by clicking the expand icon.
If an itemset $A$ is included in itemset $B$ (\ie, $A\in B$), $A$ can be opened by clicking the expand icon on $B$.
% In the main body of the attribute matrix, each discriminatory itemsets is represented by a row (\autoref{fig:interface}A3).

For each discriminatory itemset, a simple glyph (\autoref{fig:interface}(A5)) is put at the left end to indicate the degree of discrimination (\ref{goal:measure}).
The radius of the glyph encodes the size of the itemset.
The angle of the outer arc is determined by the probability that the non-protected group is labeled as beneficial class (\ie, $P(\hat{Y}=1|A=0, s)$) and the angle of the inner arc is determined by the probability that the protected group is labeled as beneficial class (\ie, $P(\hat{Y}=1|A=1, s)$).
% The color saturation of the arc is proportional to the absolute value of the risk difference.

The condition of discrimination (\ie, a series of attribute values) is represented by an array of rectangles, whose solid parts represent the specified ranges/values of the aligned attributes (\ref{goal:condition}).
\qianwen{Since we discretize continuous attributes for mining classification rules, all attributes here are actually categorical and are visualized using similar encodings with continuous attributes.}
The color saturation of these rectangles is determined by the absolute value of the risk difference and the color hue is determined by the sign of the risk difference (\ref{goal:measure}).
To better compare the condition of discrimination (\ref{goal:compare}), itemsets are vertically arranged in an order that maximizes the Jaccard similarity of adjacent itemsets.

\subsection{Interactions}

% The causal graph and the scatter plot offer users an overview of discrimination and navigate them for further exploration. 
\noindent
\textbf{Filter Itemsets \& Modify Resolving Attributes}
% In the causal graph(\autoref{fig:interface}(a)), users can modify the resolving attributes based on their domain knowledge (\ref{goal:validate}).
% To guide the setting of resolving attributes, the causal graph represents the causal relationships among attributes and highlights the candidate resolving attributes based on the results of causal modeling.
% Users can also modify resolving attributes in the head of the attribute matrix.

\noindent
The scatter plot (\autoref{fig:interface}(C)) allows the filtering of itemsets by offering a summary of all discriminatory itemsets (\ref{goal:distribution}). 
In the scatter plot, one point represents one discriminatory itemset, with the x-axis indicating the \textit{risk difference} (the extent of the unequal treatment) and the y-axis indicating the size of the itemset (\ref{goal:measure}). 
The scatter plots help users measure the discrimination from a coarse level and guide them in the filtering operation.
By moving the sliders on the x-axis, users can change the threshold of risk difference and select itemsets for further exploration.
The scatter plot also supports a coarse level comparison of models (\ref{goal:compare}) through two ways: juxtaposition (\ie, side by side) and superposition (\ie, overlay).

In the head of the attribute matrix (\autoref{fig:interface}(A1)), 
users can modify resolving attributes (\ref{goal:validate}) through drag and drop.
Such modification changes the mined discriminatory itemsets.
Other visual components, \ie, the scatter plot, the RippleSet, and the rows in the attribute matrix, will be updated dynamically.

\noindent
\textbf{Coordinate RippleSet with Attribute Matrix}

\noindent
% The attribute matrix and the RippleSet, representing discriminatory itemsets from different aspects, are coordinated to support a comprehensive analysis. 
The attribute matrix and the RippleSet are coordinated to support a comprehensive analysis. 
Each root row of the attribute matrix is linked to a RippleSet through a curve.
Clicking on a RippleSet can expand the rows belonging to the root row.
Users can hover over a row to highlight the corresponding itemset in RippleSet.
By clicking on a row, users can draw an outline for this itemset (\autoref{fig:interface}(D)) to facilitate their understanding of the discrimination distribution (\ref{goal:distribution}).

Three modes of analysis are supported: the compact mode, the parallel mode, and the comparison mode.
% The compact mode and the parallel mode help users smoothly switch between the overview and the details they are interested in. 
% Users can switch from the compact mode to the parallel mode by filtering out unimportant itemsets in the scatter plot.
In the compact mode (\autoref{fig:interface}(a)), rows in the attribute matrix are compact, and each RippleSet is placed to minimize the overall L1 distance between the RippleSets and the corresponding root rows.
The compact mode provides space efficiency and can offer an overview of a relatively large number of itemsets.
A ``focus+context'' technique is employed in the compact mode.
The focused RippleSet will be horizontally aligned with the corresponding rows (\autoref{fig:interface}(D)).
In the parallel mode (\autoref{fig:interface}(b)), RippleSets are placed vertically and each root row is horizontally aligned with the corresponding RippleSet.
The parallel mode provides clear alignment and thus can guide users on the detailed exploration of a relatively small number of itemsets.
The compact mode and the parallel mode help users smoothly switch between the overview and the details they are interested in. 
% Users can switch from the compact mode to the parallel mode by filtering out unimportant itemsets in the scatter plot.
In the comparison mode (\autoref{fig:interface}(c)), users can conduct a side-by-side comparison of the discriminatory itemsets of two machine learning models (\ref{goal:compare}).
\qianwen{
As pointed out by Marx et al.~\cite{marx2019predictive}, real-world datasets can admit ML models that have equal accuracy but assign conflicting predictions to the same input data.  
To help users identify these conflicting predictions and compare the fairness of two models, two RippleSets and their corresponding root rows are horizontally aligned if they have the same resolving attribute values.
}

%% file: sections/7.evaluation.tex
% \section{Evaluation}
% In this section, we conduct a laboratory study to assess the utility and effectiveness of DiscriLens.
% In this section, we evaluate the effectiveness of DiscriLens.
% The first task serves as active learning of

% The second task, participants were presented with individual
% story curves (without movie names) and a total of 20 multiple-choice
% questions, some of which are stated in Table 1.

% \subsection{Comparison of Set Visualizations}
% Based on the survey results in~\cite{setviz}, we compare DiscriLens with other set visualizations that can represent set intersections.
% The result is shown in \autoref{tab:set_compare}.
% DiscriLens, as a combination of Euler-based and Matrix-based set visualization, supports a more comprehensive analysis of discrimination.

\setlength\tabcolsep{3.5pt}
\begin{table}
\scriptsize{}
\caption{Tasks in the user study.}
\centering
\label{tab:user_tasks}
    \begin{tabular}{c|c|p{6.5cm}}
    \hline
         Task & Goals & Questions \\
         \hline
         
      T1 & \ref{goal:validate},\ref{goal:compare}&
      Compare the degrees of discrimination in the following settings of resolving attributes.\\
      
      T2 & \ref{goal:measure},\ref{goal:condition} &
      Which item will face the largest discrimination?\\
      
      T3 & \ref{goal:measure},\ref{goal:distribution} &
      Which itemset has a varying degree of discrimination?\\
      
      T4 & \ref{goal:condition},\ref{goal:compare} &
      Which item will be discriminated against by both model $m1$ and model $m2$?\\
      
    \bottomrule
    \end{tabular}
\end{table}

\begin{table}
\caption{Comparing different visualizations to choose a proper baseline.}
    \label{tab:set_compare}
    \centering
    \begin{tabular}{p{1.7cm}|p{0.4cm}|p{0.4cm}|p{2cm}|p{0.4cm}|p{1.8cm}}
    \toprule
          & G1 & G2 & G3 & G4 & G5 \\
         \hline
         Euler & \xmark & \cmark & implicit & \cmark & \xmark \\
         \hline
         BubbleSets & \xmark & \cmark & implicit & \cmark & \xmark \\
         \hline
         ParallelSet\newline \cite{kosara2006parallelset} & \xmark & \cmark & only categorical \newline attributes & \xmark & \xmark \\
         \hline
         UpSet\newline\cite{lex2014upset} & \xmark & \cmark & only binary \newline attributes & \cmark & only between \newline itemsets \\
         \hline
         \rowcolor{lightGray}
         DiscriLens & \cmark & \cmark & \cmark & \cmark & \cmark  \\
         \hline
    \end{tabular}
\end{table}

\begin{figure}[]
    \centering
    \includegraphics[width=\linewidth]{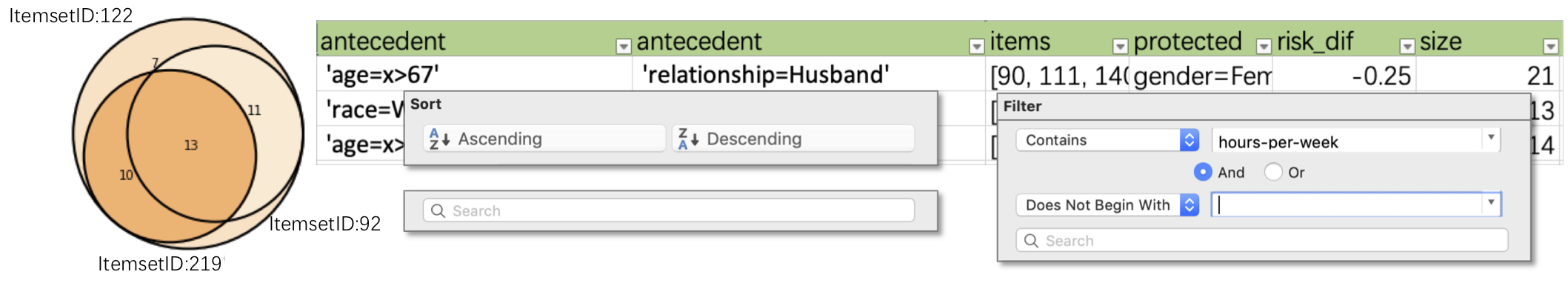}
    \vspace{-1em}
    \caption{The baseline system used in the user study: Euler diagram + table (with search, sort, and filter functions).}
    \label{fig:baseline}
\end{figure}

\begin{figure}
    \centering
    \vspace{-1em}
    \includegraphics[width=1\linewidth]{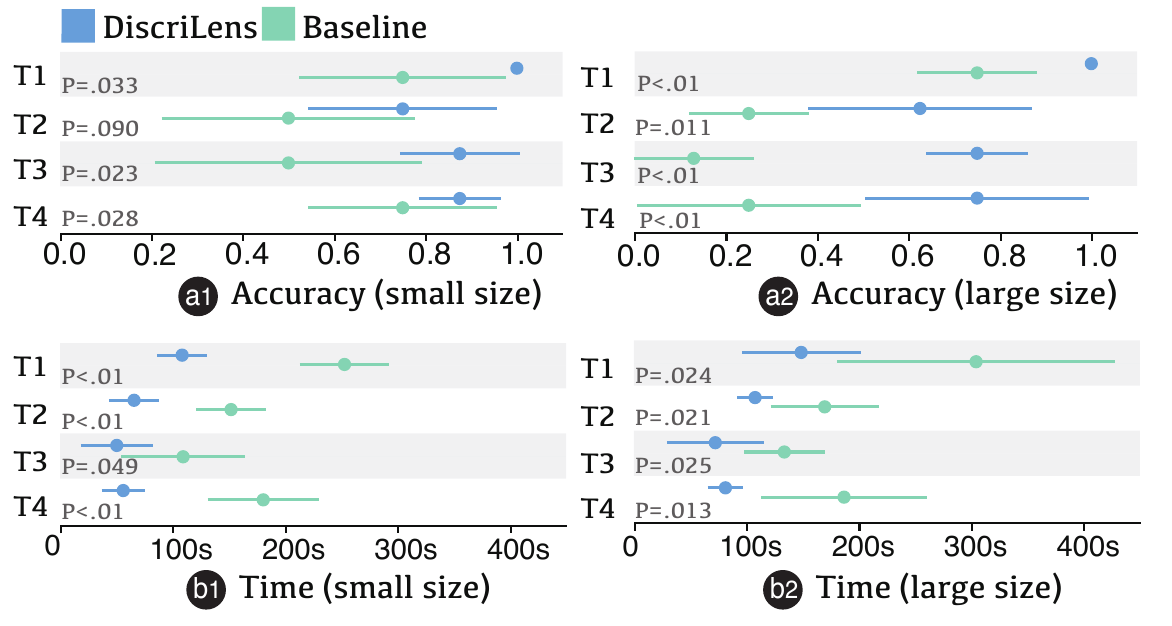}
    \vspace{-2em}
    \caption{Accuracy and completion time under different conditions. Error bars indicate 95\% confidence intervals.}
    \label{fig:time_accuracy}
\end{figure}

\section{Laboratory Study}

\qianwen{
We conducted a laboratory study to assess participants' ability to use DiscriLens to analyze machine learning discrimination.
Even though it is also important to separately evaluate RippleSet, we evaluate DiscriLens as a whole due to our focus on discrimination analysis.}
\subsection{Experimental Settings} We recruited 16 participants (12 males, 4 females, age 21-30) through university mailing lists.
All participants are postgraduate students in the school of engineering and have experiences in machine learning.

A within-subject design was employed for the user study.
We designed four tasks to assess participants' ability to conduct discrimination analysis (\textbf{G1}-\textbf{G5}), as shown in~\autoref{tab:user_tasks}. 
Each task contains 2-4 different questions in the same format.
We trained models and mined discriminatory itemsets for two datasets: an adult income dataset~\cite{uci2019} with a small number of discriminatory itemsets and a student performance dataset~\cite{amrieh2016academicDataset} with a large number of discriminatory itemsets.
\qianwen{
For the user study baseline, we first considered existing fairness visualization tools, such as FairVis~\cite{cabrera2019fairvis}, FairSight~\cite{ahn2019fairsight}, FairLearn~\cite{fairlearn}, and TensorFlow Fairness Indicator~\cite{fairnessID}.  
However, it is unfair to compare these tools with DiscriLens since they focus on different aspects and have different design goals, as discussed in~\autoref{subsec:discrimination_vis}.
}
On the contrary, we compare DiscriLens with other visualizations that can illustrate set intersections based on the survey results in~\cite{setviz}.
The result is shown in \autoref{tab:set_compare}.
DiscriLens, as a combination of Euler-based and Matrix-based set visualization, supports a more detailed analysis of discrimination.
Since no existing set visualization can meet all the requirements of discrimination analysis, we used Euler diagram + table (with search, sort, and filter functions) as the baseline method (\autoref{fig:baseline}).
% Note that we didn't choose FairVis~\cite{cabrera2019fairvis} or FairSight~\cite{ahn2019fairsight} as the baseline since they focus on different aspects and have different design goals from our tool, as discussed in~\autoref{subsec:discrimination_vis}.

For each participant, two datasets were randomly associated with the two conditions (DiscriLens and baseline) and were presented in a counter-balanced order.
\qianwen{
Before the formal study, each participant received a 20-minutes tutorial to learn the tool, complete the trial tasks, and freely ask questions.
}
In each condition, participants completed the four tasks.
\qianwen{Participants are randomly ordered in this user study.}
Finally, each participant completed a post-study questionnaire and received a short informal interview.

We set a threshold at 0.05 for p value and hypothesized that, compared with the baseline, \textbf{H1}) DiscriLens is harder to understand; \textbf{H2}) DiscriLens doesn't decrease the accuracy of tasks; \textbf{H3}) DiscriLens decreases the completion time of tasks.

\subsection{Results} 
\begin{figure}
    \centering
    \includegraphics[width=0.95\linewidth]{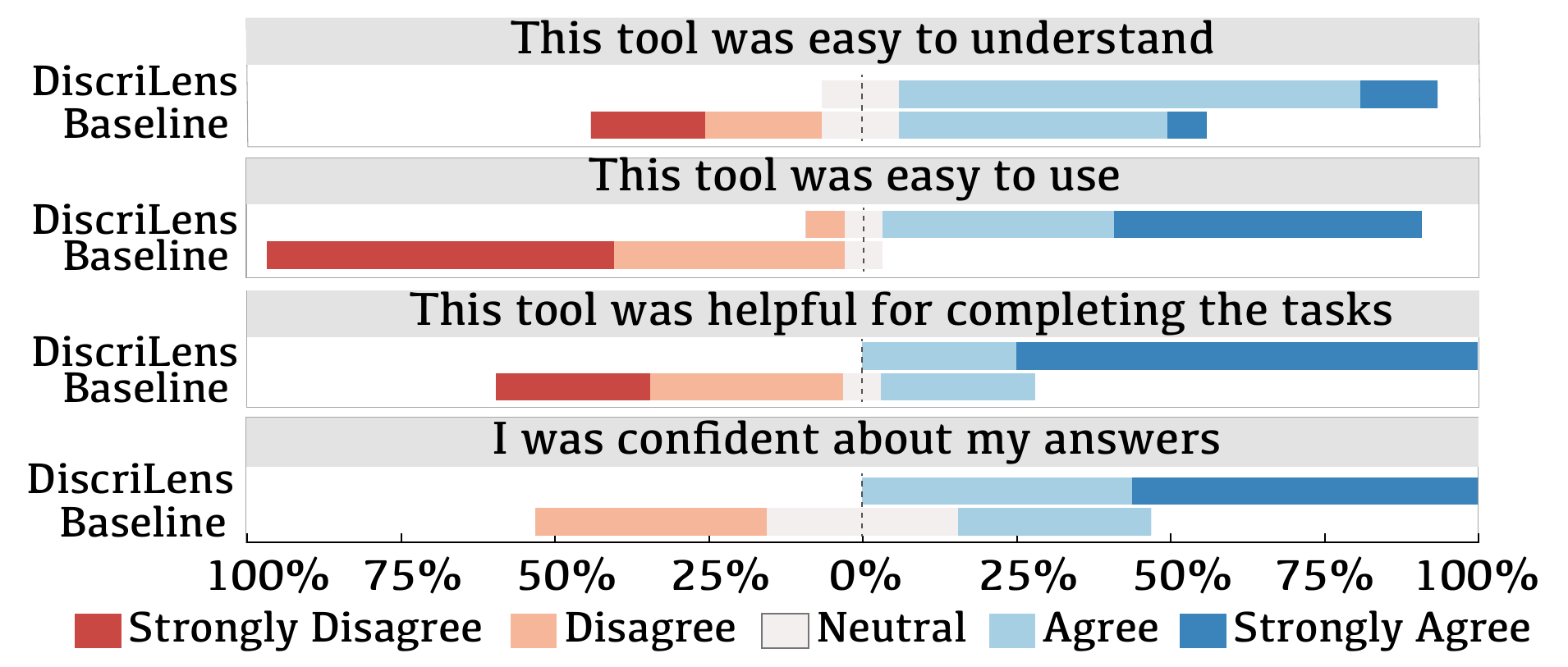}
    \vspace{-1em}
    \caption{Comparison of DiscriLens and the baseline based on the post-study questionnaires.}
    \label{fig:likert_scale}
    \vspace{-1em}
\end{figure}

% The accuracy and time cost of the performed tasks are summarized in \autoref{fig:time_accuracy}. 
% The results of the post-study questionnaire are summarized in~\autoref{fig:likert_scale}.
We summarized the results of the performed tasks in \autoref{fig:time_accuracy} and the results of the post-study questionnaire in \autoref{fig:likert_scale}.

\qianwen{
Contrary to our expectation, participants thought DiscriLens was easier to understand than the baseline (\autoref{fig:likert_scale}), rejecting \textbf{H1}. 
Although most (10/16) participants expressed that \textit{``the tool seems complicated at first glance''}, they stated that \textit{``the visual encodings are actually easy to understand and remember once you explained them''} and \textit{``the tool offers better support to the analysis than the baseline''}.
This highlighted the importance of designing visualizations suitable for the analysis tasks.
Complicated visual designs are sometimes inevitable for the analysis of complicated data. 
A simple visualization can be hard to understand when it fails to support the analysis. 
}

As shown in~\autoref{fig:time_accuracy}(a), except T2 in small size ($p=0.09$), participant gave significantly more accurate answers ($p<.05$) when using DiscriLens. This result supports \textbf{H2}. Note that the overlap between accuracy confidence intervals decreased with the increase of data size. This phenomenon might indicate that DiscriLens has a more significant advantage in accuracy when analyzing a large volume of discrimination.

As shown in~\autoref{fig:time_accuracy}(b), the time cost of DiscriLens is significantly less than that of the baseline ($p<.05$), which supports \textbf{H3}. Unlike accuracy, the overlap between time confidence intervals increased with the increase of data size. This was because participants tended to give up some tasks when they found these tasks were difficult to complete using the baseline, which led to short completion time.

Meanwhile, we observed that participants employed different strategies with the two tools. 
When using the baseline to answer a question, most participants (12/16) switched back and forth between different question options and compared them carefully using the baseline tool.
When using DiscriLens, most participants (10/16) first read the question, then examined the visualization to conclude an answer, and finally chose an option directly in the questionnaire.

\begin{figure*}
    \centering
    \includegraphics[width=\linewidth]{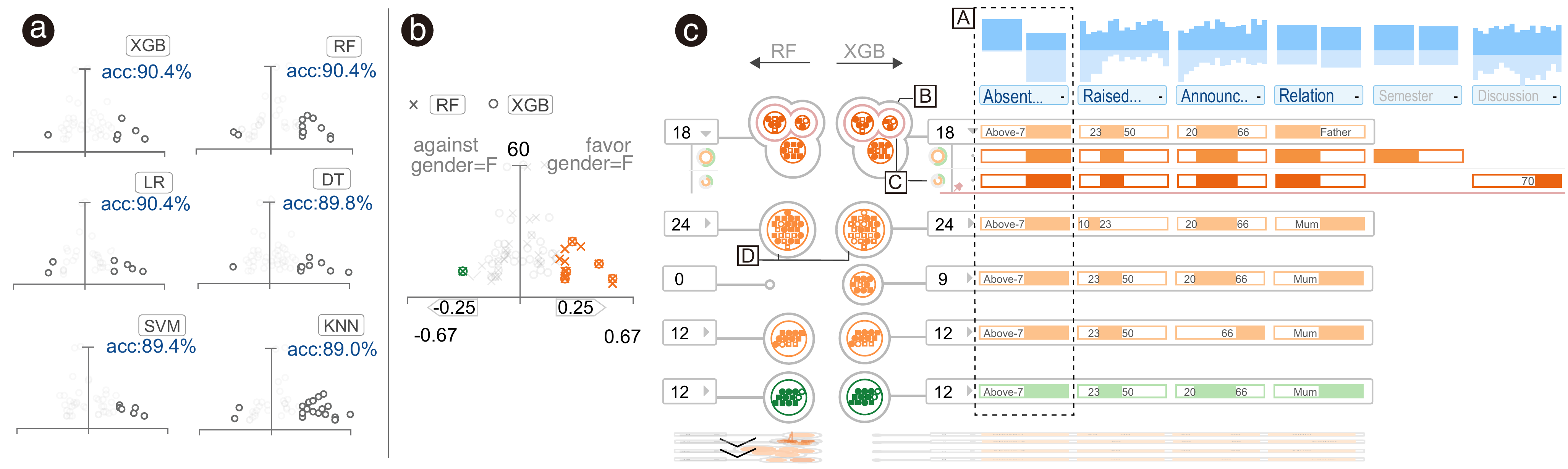}
    \vspace{-2em}
    \caption{Use DiscriLens to analyze discrimination: (a) analyze the discriminatory itemsets of different models using the scatter plots; (b) compare RF with XGBoost in a scatter plot; (c) compare RF with XGBoost in the comparison mode.}
    \label{fig:use_case}
\end{figure*}

\input{sections/case_study.tex}

%% file: sections/case_study.tex
\section{Use Cases}
\label{sec:case}

In addition to the laboratory study, we further demonstrated the effectiveness of DiscriLens in analyzing algorithmic discrimination through use cases.
% We used two publicly available datasets from different domains: a loan application dataset (Credit)~\cite{uci2019} and a student performance dataset (xAPI)~\cite{amrieh2016academicDataset}.
% The cases were conducted in collaboration with two machine learning experts (E1 and E2) and a professor with more than ten years of teaching experience (E3).
% We used two publicly available datasets from different domains: an income dataset (Adult)~\cite{uci2019} and a student performance dataset (xAPI)~\cite{amrieh2016academicDataset}.
The cases were conducted in collaboration with two machine learning experts (E1 and E2) and one domain experts: a professor with more than ten years of teaching experience (E3).

% Six different types of machine learning models were trained for both cases, \ie,
% XGBoost, k-nearest neighbor (KNN), logistic regression (LR), support vector machine (SVM), random forest (RF), and decision tree (DT).
% The xAPI Education dataset~\cite{amrieh2016academicDataset} and the Adult dataset~\cite{uci2019} were used.
We mainly used the xAPI dataset~\cite{amrieh2016academicDataset} for demonstration and more use cases are available in the supplementary material. 
Each data point in the xAPI datasets has 9 attributes (\eg, \texttt{raised hands}, \texttt{absence days}) of a student and a binary label indicating whether the test score of this student is over 69.
We set \texttt{gender=female} as the protected group and $\tau=0.25$.
% We collaborated with two machine learning experts (E1 and E2) and a professor with more than ten years of teaching experience in machine learning (E3).
Six different types of ML models were trained:
XGBoost, k-nearest neighbor (KNN), logistic regression (LR), support vector machine (SVM), random forest (RF), and decision tree (DT).
The hyperparameters of all six models have been tuned to maximize the 5-fold cross-validation accuracy using AutoML~\cite{bergstra2012random, wang2019atmseer}. More implementation details are available in the supplementary material.

% Details about the experiment settings and more use cases on other datasets were attached in the supplementary material.

% \begin{figure}[b!]
%     \centering
%     \includegraphics[width=\linewidth]{pictures/case1.pdf}
%     \caption{Use DiscriLens to analyze algorithmic discrimination: (a) set protected attribute and resolving attributes; (b) compare the discriminatory itemsets of four models and examine the details of SVM; (c) compare XGBoost with DT. }
%     \label{fig:case1}
% \end{figure}

\noindent
\textbf{Identify Discrimination.}
E3 conducted analysis on the xAPI dataset and set the resolving attributes at first. 
Based on his domain knowledge and the observation of item distribution on different attributes (\autoref{fig:interface}(A1)), 
E3 was satisfied with the resolving attributes suggested by the discovery module set \texttt{[announcements view,}\texttt{raised hands,absence days, relationship]} and did not modify them.
He then analyzed the XGBoost model, which had the highest 5-fold cross-validation accuracy.

In the training data, female and male students had a different proportion of high scores.
However, E3 observed no discriminatory itemsets in the scatter plot.
He deduced this phenomenon can be explained by the correlation between \texttt{gender} and resolving attributes. 
For example, female students might have lower values of ~\texttt{absence days}.
He then verified his conjecture by dragging \texttt{absence days} and \texttt{relationship} from the resolving attributes. Discriminatory itemsets then appeared in the scatter plot, which confirmed his conjecture.

% E3 was also curious about the model discrimination in unforeseen data.
% The joint distribution estimator in \cite{ming2018rulematrix} was used to model the distribution of training data and to generate a number of (1000 in this case) unforeseen data samples.
E3 then used the trained models to predict unforeseen samples.
As shown in~\autoref{fig:interface}, a number of discriminatory itemsets were identified and presented, indicating the model discrimination appeared in unforeseen samples.
E3 also found that the majority of the discriminatory itemsets were orange (\autoref{fig:use_case}(c)), which represented discrimination towards the non-protected group (\ie, male students).

Moreover, E2 conducted the same analysis on the Adult dataset~\cite{uci2019}. 
The Adult dataset contains 45,222 data items. The task of this dataset is to predict whether the annual income of a person is over 50k USD. 
% It contains 45,222 data items. The task of this dataset is to predict whether the annual income of a person is over 50k USD. 
Interestingly, for models trained on the Adult dataset, discrimination did not increase in test data.
One possible explanation of this phenomenon is that models trained on large datasets are more stable and make more consistent predictions. 
Another possible explanation is that the test data and training data had a very similar distribution of the Adult dataset, while the opposite goes for xAPI dataset.

\noindent
\textbf{Understand Discrimination among Itemsets.}
E2 was interested in the XGBoost result, which had the highest 5-fold cross validation accuracy. He examined the discrimination in different itemsets at XGBoost.
In the attribute matrix, the solid parts in rectangles represented the condition of discrimination. 
E2 found that all discriminatory items had the condition \texttt{absence days:>7} (\autoref{fig:use_case}A).
In other words, if a student was absent less than 7 days, this student was less likely to be treated differently based on gender. This observation can be easily explained by the histogram of \texttt{absence day}, which indicated that almost all students whose \texttt{absence day<7} were predicted as the high-score class. \looseness=-1

E2 then compared the RippleSets of different itemsets.
As shown in \autoref{fig:use_case}B, among all RippleSets, the RippleSet of \texttt{[absence days:>7, announcements view:20-66, raised hands:23-50, relationship=father]} had the highest color saturation, indicating the largest value of RD.
Meanwhile, this RippleSet consisted of three clusters of dots with different saturation, indicating that the RD value varies.
For a detailed examination, E2 clicked on this RippleSet and expanded the corresponding root row. 
% As shown in \autoref{}, two rows with varying color saturation (\ie, degree of discrimination) appeared under the root row.
As shown in \autoref{fig:use_case}(C), the row with higher saturation indicated that, inside the itemset \texttt{[absence days:>7, announcements view:20-66,raised hands: 23-50, relationship:father]}, male students whose \texttt{discussion>70} were more likely to be treated unequally. \looseness=-1

These observations reveal when the model predictions are potentially discriminatory and should be checked and modified by human experts or discrimination-removal methods.

\noindent
\textbf{Compare Discrimination among Models.}
Instead of directly applying XGBoost, the model with the highest accuracy,
E1 compared this model with other models from a fairness perspective.
According to the scatter plots (\autoref{fig:use_case}(a)),
DT and KNN had itemsets with higher values of risk difference (\ie, points with larger y-position) than XGBoost.
Therefore, these two models were first excluded.

E1 then compared XGBoost with RF, the model with the second highest accuracy.
Overall, RF had more discriminatory itemsets than XGBoost.
An interesting finding was that four of the five discriminatory itemsets in XGBoost also existed in RF:
coincident points in the scatter plot corresponded to itemsets with the same size and the same degree of discrimination~(\autoref{fig:use_case}(b)); 
aligned rows in attribute matrix indicated the same condition of discrimination~(\autoref{fig:use_case}(c)).
For these four discriminatory itemsets, the only difference was in the itemset \texttt{[absence days:>7, announcements view:20-66, raised hands:10-23, relationship:mom]]}, where the RippleSet of RF has more hollow items than XGBoost~(\autoref{fig:use_case}(D)).
In other words, XGBoost was more likely to predict students in this itemset as the low-score class.
Considering that most discriminatory itemsets in XGBoost also existed in RF and that RF had more discriminatory itemsets, E1 concluded that XGBoost was a better choice than RF even from the fairness perspective. 

E1 also compared XGBoost with LR and SVM. He found these models had discrimination towards different itemsets, as most of their RippleSets were not aligned. Therefore, comparing the discrimination of these models depended on which groups of students were more important for the model user.
E1 commented, \textit{``Without visualization, all these differences might just be obscured in a summary statistic.''} 
% Based these comparison, E1 commented that XGBoost was still a good choice for this task even when considering fairness into accout.

\begin{figure}
    \centering
    \includegraphics[width=\linewidth]{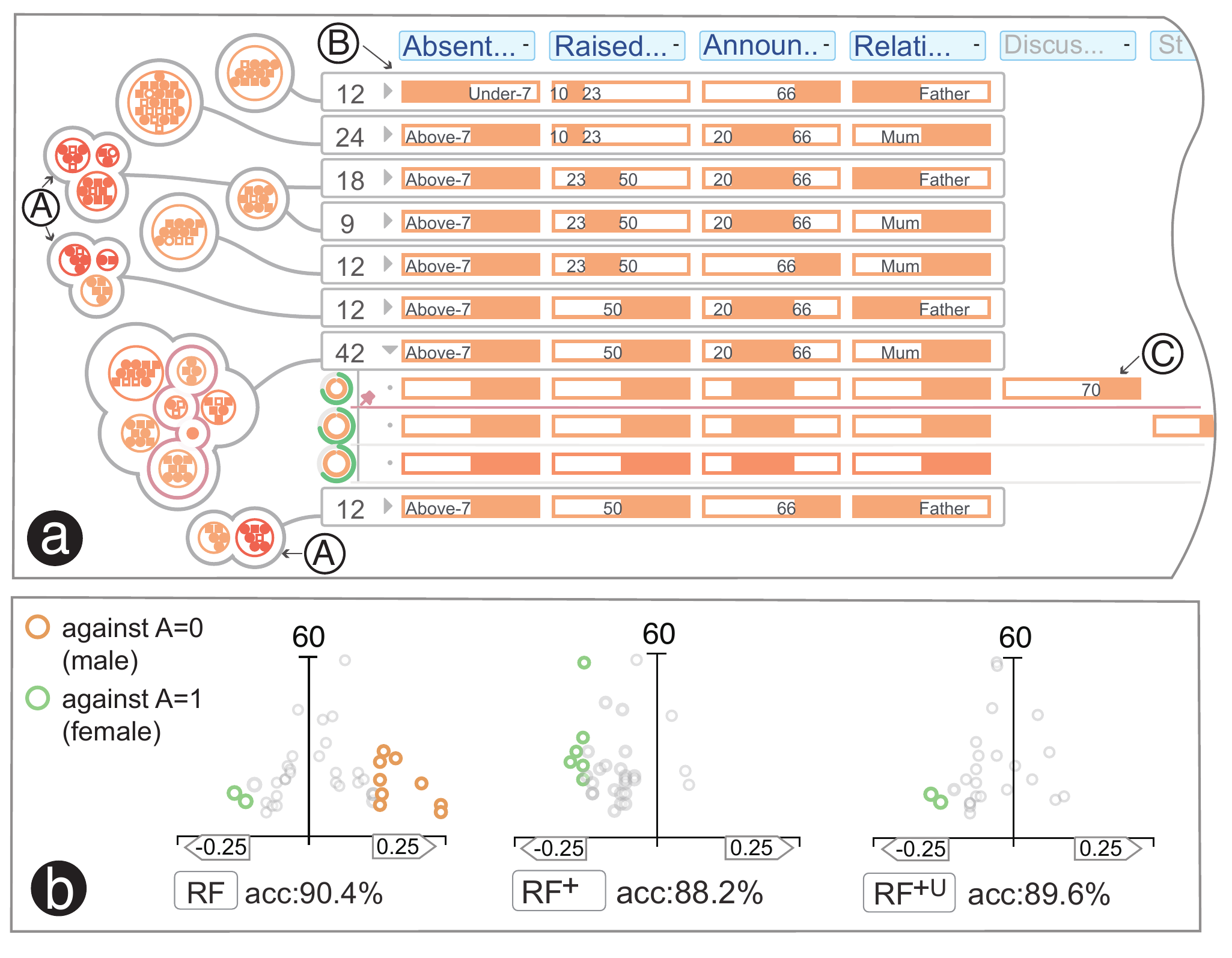}
    \vspace{-2em}
    \caption{(a) Users can identify discrimination with high values of RD (A) and discrimination towards critical groups (BC). (b) Compare the discriminatory itemsets of the original RF model, the RF model using Reject Option (RF$^+$),  and  the  RF  model  using  the  user-defined Reject Option (RF$^{+U}$).}
    \label{fig:rf_campare}
    \vspace{-1em}
\end{figure}

\noindent
\textbf{Remove Discrimination.}
E1 adopted Reject Option~\cite{kamiran2012decision}, a popular discrimination-removal algorithm~\footnote{The implementation by IBM Fairness 360~\cite{aif360} is used here.}, to remove discrimination in RF, a model with relatively more complex discriminatory itemsets.

One solution is to apply Reject Option to all discriminatory itemsets whose $RD>\tau=0.25$.
However, like most discrimination-removal methods~\cite{kusner2017counterfactual, speicher2018unified}, this operation led to unexpected reverse discrimination (\ie, towards female students) and a decrease in accuracy.
% To solve this problem, 
One possible solution to this problem is to increase $\tau$ and apply Reject Option to discrimination with higher RD values.

However, in practice, severe discrimination is not only those with high RD values but also those towards critical groups. 
For example, E3 was concerned with the discrimination happened to hard-working students (\eg, low absenteeism, active discussion).  
Therefore, E1 employed DiscriLens to identify severe discrimination.
Discriminatory itemsets with high RD can be easily identified through their colors (\autoref{fig:rf_campare}(A)).
E1 then checked the discrimination towards critical groups. 
He first examined the students with low absenteeism and selected the itemset whose \texttt{absent days<7} (\autoref{fig:rf_campare}(B)). 
He did not open the row for further examination since the corresponding RippeleSet, with only one cluster of items, indicated there was no complex intersection. 
E1 then checked the students with active class participation by examining the values of \texttt{raised hands}.
Among the three RippleSets with high values ($>50$) of \texttt{raised hands}, two had high values of RD and only one~(\autoref{fig:rf_campare}(C)) was left to be examined. 
E1 expanded the rows of this RippleSet and selected an itemset~(\autoref{fig:interface}(F)) that also had a high value \qianwen{($>70$)} of discussion.

Reject Option was then applied to the user-defined discriminatory region.
For convenience, we denoted RF using default discriminatory region by RF$^{+}$, and RF using the user-defined discriminatory region by RF$^{+U}$.
As shown in ~\autoref{fig:rf_campare}(b),
compared with RF$^+$, RF$^{+U}$ reduced both the accuracy loss and the reverse discrimination \mycircle. 
By providing more detailed information, DiscriLens enables users a precise removal of discrimination.

\noindent
\textbf{Expert Interview}
All the experts expressed great enthusiasm for DiscriLens and offered useful suggestions for its improvement.
They commented that it helped them \textit{``understand ML models from a novel perspective''} (E3) and could \textit{``contribute to the adoption of ML to real-world applications''} (E1).
E1 suggested combining the analysis of model accuracy with the analysis of discrimination.
E2 was curious about the comparison between algorithmic discrimination and human discrimination, \textit{``Even though a model makes discriminatory predictions, it can still be less biased than humans.''}
E3 suggested adding a \textit{``what-if''} function to check whether and how a model would discriminate against a given data item.

The issue of trust was mentioned and discussed by the experts.
% E3 asked: \textit{``How to ensure all attributes related to the student performance have been included in the dataset?''}
E3 suspected that the training data might omit some attributes related to the students performance.
He also commented that the results of discrimination analysis were influenced by the choices of the analysts (\eg, the setting of key attributes).
\textit{``Two analysts might draw opposite conclusions.''}
% E4 stated: \textit{``I feel more reliable about the analysis based on ground truth than the analysis based on the comparison between two groups.''}.
% But he understood \textit{``it is difficult, or even impossible, to obtain ground truth of discrimination in real applications''}.
E1, the machine learning expert, suggested the comparison between algorithmic discrimination and human discrimination to increase user confidence. 
\textit{``Even though a model makes discriminatory predictions, it can still be less biased than humans and thus be helpful.''}

% The issue of trust was mentioned by the experts.
% E3 suspected that the training data might omit some attributes related to the students performance.
% E4 commented that the results of discrimination analysis were influenced by the choices of the analysts (\eg, the setting of key attributes).
% \textit{``Two analysts might draw different conclusions.''}(E3)

%% file: sections/8.discussion_conclusion.tex
\section{Discussion}

% \subsection{Generality}
% In this paper, we focus on binary classifiers for the analysis of algorithmic discrimination.
% However, the proposed method can be applied to multi-class scenarios by decomposing a multi-class classification into a series of one-vs-remaining classification.
% Meanwhile, even though DiscriLens is designed for analyzing algorithmic discrimination, it can be easily extended to unveil human discriminatory practices in the historical decision records (\eg, college admission records) and help prepare high-quality data for the model training.
% Moreover, the novel set visualization we propose in DiscriLens can be applied to analyzing other types of set data.
% % can be used to analyzed the training data; similar to the methods discussed in ~\cite{luong2011k, pedreshi2008discrimination}
% % investigate discrimination with various data distribution
% % generalized to multi-class problem

\subsection{Scalability}
For the discrimination discovery, it takes about six minutes to run the four-stage pipeline on 4,000 samples with 14 features on a PC (2.3GHz dual-core, Intel Core i5 processor).
The major bottlenecks lie in the FEGS algorithm (two minutes), the FP-Growth algorithm (one minute), and the discriminatory rules mining (three minutes).

The scalability of Rippleset is mainly limited by the number of sets and the number of items. According to our user study, Rippleset effectively reduces the visual clutter in traditional Euler diagrams, which has difficulties to handle more than three sets. 
However, the readability of RippleSet hasn't been validated with more than seven intertwining sets.
\qianwen{For a large number of items, RippleSet supports the aggregation of items, as shown in~\autoref{fig:interface}(D1). 
In the future, we plan to support more grouping strategies, such as those discussed in Squares~\cite{ren2017squares}, and provide more statistical summaries in the aggregation view of RippleSet.
}

% % For the discrimination visualization, its effectiveness is limited by the number of attributes and the number of itemsets.
% Meanwhile, the number of attributes may also influences the effectiveness of the visual design.
% The widely-used datasets in algorithmic discrimination, such as Adult, Credit Approval, and Bank Marketing~\cite{uci2019}, have no more than 15 features and no more than five key attributes. 
% The proposed visualization can be safely applied to these common datasets.
% % Therefore, we assume the proposed visualization is effective in common cases.

\qianwen{
\vspace{-4mm}
\subsection{Learning Curve}
We acknowledge that the novel and complicated visualization designs in DiscriLens can pose challenges to users, especially those with no prior knowledge in visual analytics. 
Even though we have provided interactive tutorials and legends to alleviate this issue, the steep learning curve still exists and can lead to low accessibility to novices. 
For example, RippleSet and the attribute matrix can be overwhelming when there are too many itemsets. 
Users may stop at the scatter plot, blindly trying to optimize RD without conducting detailed visual analysis of discrimination.
% Users need to spend time learning the visualization design before they can effectively conduct discrimination analysis using DiscriLens. 
On the other hand, complicated visual designs are sometimes inevitable for the analysis of complicated data~\cite{wang2018narvis, chen2007handbook}. 
Even with easy-to-understand encodings, a simple visualization can be hard to use when it fails to support the required analysis.
How to strike a good balance between designing intuitive visualizations and accomplishing complex analysis tasks is still an open question in the field of visual analytics and requires further investigation.
}

\qianwen{
\subsection{Evaluation of RippleSet}
In this paper, we proposed a novel set visualization, RippleSet, to assist the analysis of discrimination.
RippleSet supplements the attribute matrix by illustrating the discrimination distribution among data items, especially these belong to intricately intertwined sets.
While the current evaluation shows the utility of DiscriLens as a whole, it fails to demonstrate the contribution of RippleSet to the effectiveness of DiscriLens.
Apart from discrimination analysis, RippleSet also shows the potential to stand alone and be applied for more general tasks.
Unfortunately, due to page limits and our focus on discrimination analysis, we are unable to provide a stand-alone evaluation of RippleSet in this paper.
}

\subsection{Subjectivity in Analysis}
In DiscriLens, we allow users to customize the definition of discrimination and support the integration of human domain knowledge.
While this feature is regarded as a strength by the interviewed experts, we also admit that user customization can act as a double-edged sword in discrimination analysis.

On the one hand, user customization enables the integration of domain knowledge. Since the definition of discrimination differs across domains, the involvement of domain knowledge enables a more comprehensive analysis. 
On the other hand, additional bias can be potentially imposed by the subjective choices of the analysts. 
Different customization may lead to different analysis results.
One possible solution is to support what-if analysis and cross-validate the analysis results of different customization.
Analysts can test and compare different customization before drawing the final conclusion.
Similar methods are commonly used in verifying subjective analysis.

\subsection{More Intelligent Discrimination Mining}
\qianwen{
In this work, we do not consider the hidden attributes in the causal graph(\ie, attributes that cannot be explicitly observed).
We assume the key attributes can capture the main model decisions and set a threshold of the risk difference to allow oscillations caused by hidden attributes.
However, the protected or resolving attributes may not be included in the dataset, and this assumption sometimes can fail.
In future work, we plan to take the hidden attributes into consideration and offer more comprehensive explanations of the model predictions.
}

\qianwen{Meanwhile, the current version of DiscriLens only supports the analysis of one protected attribute and requires users to define the protected group as input.
When there are many groups (\eg, females, blacks, LGBTs) qualified for special protection by a law, users cannot analyze multiple protected attributes at the same time and need to examine different attributes separately one by one.
% In future work, we plan to support the analysis of multiple protected attributes.
% In future work, we plan to facilitate the setting of protected group by automatically checking discrimination based on common protected attributes, such as the nine protected attributes defined in Equality Act 2010~\cite{uk-equal}.
}

\qianwen{
Compared with other explainable models (\eg, decision tree), our study provides limited support in explaining the identified discrimination.
Therefore, an interesting future direction is to integrate discrimination analysis with model interpretation and diagnosis techniques, which can help us track the origin of algorithmic discrimination and better understand model behaviors.
% For example, we can employ these techniques to understand why ensemble models increase discrimination towards certain itemsets.
% We can use these techniques to track the origin of algorithmic discrimination and better understand the diverging discriminatory behaviors of models trained on the same dataset.
}

% An interesting direction is to integrate discrimination analysis with model interpretation and diagnosis techniques.
% % For example, we can employ these techniques to understand why ensemble models increase discrimination towards certain itemsets.
% We can use these techniques to track the origin of algorithmic discrimination and better understand the diverging discriminatory behaviors of models trained on the same dataset.

\section{Conclusion}
In this work, we designed and developed DiscriLens, an interactive visualization tool that facilitated a better understanding and analysis of algorithmic discrimination.
A four-stage pipeline was developed for the discovery of discriminatory predictions.
For an effective presentation, a novel set visualization was designed by combining an extended Euler diagram with a matrix-based set visualization. 
Two case studies demonstrated the usability and utility of DiscriLens in understanding and removing algorithmic discrimination.
Context-aware Reject Option, a post-processing method, was proposed to better remove discrimination while reducing the accuracy loss.
We also reported insights into algorithmic discrimination that were obtained during the development and evaluation of DiscriLens. 

%% file: main.bbl
\begin{thebibliography}{10}

\bibitem{ahn2019fairsight}
Y.~Ahn and Y.-R. Lin.
\newblock Fairsight: Visual analytics for fairness in decision making.
\newblock {\em arXiv preprint arXiv:1908.00176}, 2019.

\bibitem{alper2011linesets}
B.~Alper, N.~Riche, G.~Ramos, and M.~Czerwinski.
\newblock Design study of linesets, a novel set visualization technique.
\newblock {\em IEEE Transactions on Visualization and Computer Graphics},
  17(12):2259--2267, 2011.

\bibitem{Alsallakh2013radicalset}
B.~Alsallakh, W.~Aigner, S.~Miksch, and H.~Hauser.
\newblock Radial sets: Interactive visual analysis of large overlapping sets.
\newblock {\em IEEE Transactions on Visualization and Computer Graphics},
  19(12):2496--2505, 2013.

\bibitem{alsallakh2014setsurvey}
B.~Alsallakh, L.~Micallef, W.~Aigner, H.~Hauser, S.~Miksch, and P.~Rodgers.
\newblock Visualizing sets and set-typed data: State-of-the-art and future
  challenges.
\newblock In {\em Eurographics Conference on Visualization, EuroVis}, pp.
  1--21. Eurographics Association, Swansea, UK, 2014.

\bibitem{alsallakh2017powerset}
B.~Alsallakh and L.~Ren.
\newblock Powerset: A comprehensive visualization of set intersections.
\newblock {\em IEEE Transactions on Visualization and Computer Graphics},
  23(1):361--370, 2017.

\bibitem{amrieh2016academicDataset}
E.~A. Amrieh, T.~Hamtini, and I.~Aljarah.
\newblock Mining educational data to predict student’s academic performance
  using ensemble methods.
\newblock {\em International Journal of Database Theory and Application},
  9(8):119--136, 2016.

\bibitem{bergstra2012random}
J.~Bergstra and Y.~Bengio.
\newblock Random search for hyper-parameter optimization.
\newblock {\em Journal of Machine Learning Research}, 13(Feb):281--305, 2012.

\bibitem{attacking-discrimination-in-ml}
G.~{B}ig {P}icture.
\newblock Attacking discrimination in {ML}.
\newblock
  \url{http://research.google.com/bigpicture/attacking-discrimination-in-ml/},
  2017.
\newblock Accessed: 2019-06-30.

\bibitem{cabrera2019fairvis}
{\'A}.~A. Cabrera, W.~Epperson, F.~Hohman, M.~Kahng, J.~Morgenstern, and D.~H.
  Chau.
\newblock Fairvis: Visual analytics for discovering intersectional bias in
  machine learning.
\newblock {\em arXiv preprint arXiv:1904.05419}, 2019.

\bibitem{chen2020}
C.~{Chen}, J.~{Yuan}, Y.~{Lu}, Y.~{Liu}, H.~{Su}, S.~{Yuan}, and S.~{Liu}.
\newblock Oodanalyzer: Interactive analysis of out-of-distribution samples.
\newblock {\em IEEE Transactions on Visualization and Computer Graphics}, 2020.
  to be published. doi: {{%
10\hspace{.1pt}\discretionary{.}{%
}{.}\hspace{.4pt}1109\discretionary{/}{%
}{/}TVCG\hspace{.1pt}\discretionary{.}{%
}{.}\hspace{.4pt}2020\hspace{.1pt}\discretionary{.}{%
}{.}\hspace{.4pt}2973258}}


\bibitem{chen2007handbook}
C.-h. Chen, W.~K. H{\"a}rdle, and A.~Unwin.
\newblock {\em Handbook of data visualization}.
\newblock Springer Science \& Business Media, 2007.

\bibitem{chickering2002bayesSearch}
D.~M. Chickering.
\newblock Optimal structure identification with greedy search.
\newblock {\em Journal of Machine Learning Research}, 3(Nov):507--554, 2002.

\bibitem{collins2009bubbleset}
C.~Collins, G.~Penn, and S.~Carpendale.
\newblock Bubble sets: Revealing set relations with isocontours over existing
  visualizations.
\newblock {\em IEEE Transactions on Visualization and Computer Graphics},
  15(6):1009--1016, 2009.

\bibitem{corbett2017algorithmic}
S.~Corbett-Davies, E.~Pierson, A.~Feller, S.~Goel, and A.~Huq.
\newblock Algorithmic decision making and the cost of fairness.
\newblock In {\em Proceedings of the 23rd ACM SIGKDD International Conference
  on Knowledge Discovery and Data Mining}, pp. 797--806. {ACM}, {Halifax, NS,
  Canada}, 2017.

\bibitem{eu-equal}
{Council of European Union}.
\newblock {Equal Employment Opportunity Commission}, 1978.
\newblock
  \newline\url{https://eur-lex.europa.eu/legal-content/EN/ALL/?uri=CELEX:32006L0054}.

\bibitem{demiralp2014learning}
{\c{C}}.~Demiralp, M.~S. Bernstein, and J.~Heer.
\newblock Learning perceptual kernels for visualization design.
\newblock {\em IEEE Transactions on Visualization and Computer Graphics},
  20(12):1933--1942, 2014.

\bibitem{Dinkla2012kelp}
K.~Dinkla, M.~J. van Kreveld, B.~Speckmann, and M.~A. Westenberg.
\newblock Kelp diagrams: Point set membership visualization.
\newblock {\em Computer Graphics Forum}, 31(3pt1):875--884, 2012.

\bibitem{uci2019}
D.~Dua and C.~Graff.
\newblock {UCI} machine learning repository, 2017.

\bibitem{Dwork2012fairness}
C.~Dwork, M.~Hardt, T.~Pitassi, O.~Reingold, and R.~Zemel.
\newblock Fairness through awareness.
\newblock In {\em Proceedings of the 3rd Innovations in Theoretical Computer
  Science Conference}, ITCS '12, pp. 214--226. ACM, New York, NY, USA, 2012.

\bibitem{fayyad1993multi}
U.~Fayyad and K.~Irani.
\newblock Multi-interval discretization of continuous-valued attributes for
  classification learning.
\newblock In {\em 13th International Joint Conference on Artificial
  Intelligence}, vol.~2, pp. 1022--1027. Morgan Kaufmann, {Chamb{\'{e}}ry,
  France}, 1993.

\bibitem{freiler2008interactive}
W.~Freiler, K.~Matkovic, and H.~Hauser.
\newblock Interactive visual analysis of set-typed data.
\newblock {\em IEEE Transactions on Visualization and Computer Graphics},
  14(6):1340--1347, 2008.

\bibitem{gerber2014predictingCrime}
M.~S. Gerber.
\newblock Predicting crime using twitter and kernel density estimation.
\newblock {\em Decision Support Systems}, 61:115--125, 2014.

\bibitem{Hajian2016AB}
S.~Hajian, F.~Bonchi, and C.~Castillo.
\newblock Algorithmic bias: From discrimination discovery to fairness-aware
  data mining.
\newblock In {\em Proceedings of the 22nd ACM SIGKDD International Conference
  on Knowledge Discovery and Data Mining}, pp. 2125--2126. ACM, {San Francisco,
  CA, USA}, 2016.

\bibitem{han2000fp-growth}
J.~Han and J.~Pei.
\newblock Mining frequent patterns by pattern-growth: methodology and
  implications.
\newblock {\em ACM SIGKDD Explorations Newsletter}, 2(2):14--20, 2000.

\bibitem{hardt2016equality}
M.~Hardt, E.~Price, and N.~Srebro.
\newblock Equality of opportunity in supervised learning.
\newblock In {\em Advances in Neural Information Processing Systems}, pp.
  3315--3323. Curran Associates, {Barcelona, Spain}, 2016.

\bibitem{holstein2019improving}
K.~Holstein, J.~Wortman~Vaughan, H.~Daum{\'e}~III, M.~Dudik, and H.~Wallach.
\newblock Improving fairness in machine learning systems: What do industry
  practitioners need?
\newblock In {\em Proceedings of the 2019 CHI Conference on Human Factors in
  Computing Systems}, p. 600. ACM, Glasgow, UK, 2019.

\bibitem{aif360}
IBM.
\newblock {AI} fairness 360.
\newblock \url{https://aif360.mybluemix.net/}, 2018.
\newblock Accessed: 2019-06-30.

\bibitem{SimpsonParadox}
P.~H.~G. Ivan~Koswara, Ananya~Aaniya.
\newblock Simpson's paradox.
\newblock \url{http://https://brilliant.org/wiki/simpsons-paradox/}.
\newblock Accessed: 2019-12-30.

\bibitem{kamiran2012data}
F.~Kamiran and T.~Calders.
\newblock Data preprocessing techniques for classification without
  discrimination.
\newblock {\em Knowledge and Information Systems}, 33(1):1--33, 2012.

\bibitem{kamiran2012decision}
F.~Kamiran, A.~Karim, and X.~Zhang.
\newblock Decision theory for discrimination-aware classification.
\newblock In {\em IEEE 12th International Conference on Data Mining}, pp.
  924--929. IEEE, Brussels, 2012.

\bibitem{kamishima2012fairness}
T.~Kamishima, S.~Akaho, H.~Asoh, and J.~Sakuma.
\newblock Fairness-aware classifier with prejudice remover regularizer.
\newblock In {\em Joint European Conference on Machine Learning and Knowledge
  Discovery in Databases}, pp. 35--50. Springer, {Bristol, UK}, 2012.

\bibitem{setviz}
Keshif.
\newblock Visual techniques for analyzing set-typed data.
\newblock \url{https://gallery.keshif.me/setvis}, 2010.
\newblock Accessed: 2019-07-30.

\bibitem{khandani2010consumer}
A.~E. Khandani, A.~J. Kim, and A.~W. Lo.
\newblock Consumer credit-risk models via machine-learning algorithms.
\newblock {\em Journal of Banking and Finance}, 34(11):2767--2787, 2010.

\bibitem{Kilbertus2017avoiding}
N.~Kilbertus, M.~Rojas-Carulla, G.~Parascandolo, M.~Hardt, D.~Janzing, and
  B.~Sch\"{o}lkopf.
\newblock Avoiding discrimination through causal reasoning.
\newblock In {\em Proceedings of the 31st International Conference on Neural
  Information Processing Systems}, pp. 656--666. Curran Associates, Long Beach,
  California, USA, 2017.

\bibitem{Kim2007conset}
B.~Kim, B.~Lee, and J.~Seo.
\newblock Visualizing set concordance with permutation matrices and fan
  diagrams.
\newblock {\em Interacting with Computers}, 19(5-6):630--643, 2007.

\bibitem{kosara2006parallelset}
R.~Kosara, F.~Bendix, and H.~Hauser.
\newblock Parallel sets: Interactive exploration and visual analysis of
  categorical data.
\newblock {\em IEEE Transactions on Visualization and Computer Graphics},
  12(4):558--568, 2006.

\bibitem{kusner2017counterfactual}
M.~J. Kusner, J.~Loftus, C.~Russell, and R.~Silva.
\newblock Counterfactual fairness.
\newblock In {\em Advances in Neural Information Processing Systems}, pp.
  4066--4076. Curran Associates, Long Beach, CA, {USA}, 2017.

\bibitem{fairlearn}
F.~Learn.
\newblock Fair learn.
\newblock \url{https://github.com/fairlearn/fairlearn}, 2019.

\bibitem{lex2014upset}
A.~Lex, N.~Gehlenborg, H.~Strobelt, R.~Vuillemot, and H.~Pfister.
\newblock Upset: visualization of intersecting sets.
\newblock {\em IEEE Transactions on Visualization and Computer Graphics},
  20(12):1983--1992, 2014.

\bibitem{li2009evaluation}
J.~Li, J.~J. van Wijk, and J.-B. Martens.
\newblock Evaluation of symbol contrast in scatterplots.
\newblock In {\em 2009 IEEE Pacific Visualization Symposium}, pp. 97--104.
  IEEE, 2009.

\bibitem{liu1997metaphor}
C.~H. Liu.
\newblock Form symbolism, analogy, and metaphor.
\newblock {\em Psychonomic Bulletin \& Review}, 4(4):546--551, 1997.

\bibitem{liu2017towards}
M.~Liu, J.~Shi, Z.~Li, C.~Li, J.~Zhu, and S.~Liu.
\newblock Towards better analysis of deep convolutional neural networks.
\newblock {\em IEEE Transactions on Visualization and Computer Graphics},
  23(1):91--100, 2017.

\bibitem{luong2011k}
B.~T. Luong, S.~Ruggieri, and F.~Turini.
\newblock k-nn as an implementation of situation testing for discrimination
  discovery and prevention.
\newblock In {\em Proceedings of the 17th ACM SIGKDD International Conference
  on Knowledge Discovery and Data Mining}, pp. 502--510. ACM, San Diego, USA,
  2011.

\bibitem{mancuhan2014combating}
K.~Mancuhan and C.~Clifton.
\newblock Combating discrimination using bayesian networks.
\newblock {\em Artificial intelligence and law}, 22(2):211--238, 2014.

\bibitem{marx2019predictive}
C.~T. Marx, F.~d.~P. Calmon, and B.~Ustun.
\newblock Predictive multiplicity in classification.
\newblock {\em arXiv preprint arXiv:1909.06677}, 2019.

\bibitem{ming2018rulematrix}
Y.~Ming, H.~Qu, and E.~Bertini.
\newblock Rulematrix: visualizing and understanding classifiers with rules.
\newblock {\em IEEE transactions on visualization and computer graphics},
  25(1):342--352, 2018.

\bibitem{naff1995subjective}
K.~C. Naff.
\newblock Subjective vs. objective discrimination in government: Adding to the
  picture of barriers to the advancement of women.
\newblock {\em Political Research Quarterly}, 48(3):535--557, 1995.

\bibitem{uk-sex}
{Parliament of the United Kingdom}.
\newblock {Sex Discrimination Act 1975}, 1975.
\newblock \newline\url{https://www.legislation.gov.uk/ukpga/1975/65}.

\bibitem{uk-equal}
{Parliament of the United Kingdom}.
\newblock {Equality Act 2010}, 2010.
\newblock \newline\url{http://www.legislation.gov.uk/ukpga/2010/15}.

\bibitem{pedreschi2009measuring}
D.~Pedreschi, S.~Ruggieri, and F.~Turini.
\newblock Measuring discrimination in socially-sensitive decision records.
\newblock In {\em Proceedings of the 2009 SIAM International Conference on Data
  Mining}, pp. 581--592. SIAM, Nevada, US, 2009.

\bibitem{pedreschi2012study}
D.~Pedreschi, S.~Ruggieri, and F.~Turini.
\newblock A study of top-k measures for discrimination discovery.
\newblock In {\em Proceedings of the 27th Annual ACM Symposium on Applied
  Computing}, pp. 126--131. ACM, Riva, Italy, 2012.

\bibitem{pedreshi2008discrimination}
D.~Pedreshi, S.~Ruggieri, and F.~Turini.
\newblock Discrimination-aware data mining.
\newblock In {\em Proceedings of the 14th ACM SIGKDD International Conference
  on Knowledge Discovery and Data Mining}, pp. 560--568. ACM, Las Vegas, USA,
  2008.

\bibitem{ragab2014comparative}
A.~H.~M. Ragab, A.~Y. Noaman, A.~S. Al-Ghamdi, and A.~I. Madbouly.
\newblock A comparative analysis of classification algorithms for students
  college enrollment approval using data mining.
\newblock In {\em Proceedings of the 2014 Workshop on Interaction Design in
  Educational Environments}, p. 106. ACM, Albacete, Spain, 2014.

\bibitem{ren2017squares}
D.~Ren, S.~Amershi, B.~Lee, J.~Suh, and J.~D. Williams.
\newblock Squares: Supporting interactive performance analysis for multiclass
  classifiers.
\newblock {\em IEEE Transactions on Visualization and Computer Graphics},
  23(1):61--70, 2017.

\bibitem{riche2010untangling}
N.~H. Riche and T.~Dwyer.
\newblock Untangling euler diagrams.
\newblock {\em IEEE Transactions on Visualization and Computer Graphics},
  16(6):1090--1099, 2010.

\bibitem{romei2014multidisciplinary}
A.~Romei and S.~Ruggieri.
\newblock A multidisciplinary survey on discrimination analysis.
\newblock {\em The Knowledge Engineering Review}, 29(5):582--638, 2014.

\bibitem{simonetto2009fully}
P.~Simonetto, D.~Auber, and D.~Archambault.
\newblock Fully automatic visualisation of overlapping sets.
\newblock {\em Computer Graphics Forum}, 28(3):967--974, 2009.

\bibitem{speicher2018unified}
T.~Speicher, H.~Heidari, N.~Grgic-Hlaca, K.~P. Gummadi, A.~Singla, A.~Weller,
  and M.~B. Zafar.
\newblock A unified approach to quantifying algorithmic unfairness: Measuring
  individual and group unfairness via inequality indices.
\newblock In {\em Proceedings of the 24th ACM SIGKDD International Conference
  on Knowledge Discovery and Data Mining}, pp. 2239--2248. ACM, London, UK,
  2018.

\bibitem{fairnessID}
TensorFlow.
\newblock Fairness indicators.
\newblock \url{https://github.com/tensorflow/fairness-indicators}, 2019.

\bibitem{us-employ}
{The United State}.
\newblock {Equal Employment Opportunity Commission}, 2010.
\newblock \newline\url{https://www.law.cornell.edu/cfr/text/29/1607.4}.

\bibitem{us-equal}
{U.S. Federal Legislation}.
\newblock {Civil Rights Act of 1991}, 1991.
\newblock
  \newline\url{https://www.eeoc.gov/eeoc/history/35th/1990s/civilrights.html}.

\bibitem{wang2018narvis}
Q.~Wang, Z.~Li, S.~Fu, W.~Cui, and H.~Qu.
\newblock Narvis: Authoring narrative slideshows for introducing data
  visualization designs.
\newblock {\em IEEE transactions on visualization and computer graphics},
  25(1):779--788, 2018.

\bibitem{wang2019atmseer}
Q.~Wang, Y.~Ming, Z.~Jin, Q.~Shen, D.~Liu, M.~J. Smith, K.~Veeramachaneni, and
  H.~Qu.
\newblock Atmseer: Increasing transparency and controllability in automated
  machine learning.
\newblock In {\em Proceedings of the 2019 Conference on Human Factors in
  Computing Systems, {CHI}}, p. 681. {ACM}, Glasgow, Scotland, UK, 2019.

\bibitem{wang2006visualization}
W.~Wang, H.~Wang, G.~Dai, and H.~Wang.
\newblock Visualization of large hierarchical data by circle packing.
\newblock In {\em Proceedings of the SIGCHI Conference on Human Factors in
  Computing Systems}, pp. 517--520. ACM, Montr{\'{e}}al, Qu{\'{e}}bec, Canada,
  2006.

\bibitem{wu2010opinionseer}
Y.~Wu, F.~Wei, S.~Liu, N.~Au, W.~Cui, H.~Zhou, and H.~Qu.
\newblock Opinionseer: interactive visualization of hotel customer feedback.
\newblock {\em IEEE Transactions on Visualization and Computer Graphics},
  16(6):1109--1118, 2010.

\bibitem{yin2003cpar}
X.~Yin and J.~Han.
\newblock Cpar: Classification based on predictive association rules.
\newblock In {\em Proceedings of the 2003 SIAM International Conference on Data
  Mining}, pp. 331--335. SIAM, San Francisco, CA, USA, 2003.

\bibitem{yuan2021survey}
J.~Yuan, C.~Chen, W.~Yang, M.~Liu, J.~Xia, and S.~Liu.
\newblock A survey of visual analytics techniques for machine learning.
\newblock {\em Computational Visual Media}, 7(1):1--31, 2021.

\bibitem{zafar2015fairness}
M.~B. Zafar, I.~Valera, M.~G. Rodriguez, and K.~P. Gummadi.
\newblock Fairness constraints: Mechanisms for fair classification.
\newblock {\em arXiv preprint arXiv:1507.05259}, 2015.

\bibitem{zhang2017achievingData}
L.~Zhang, Y.~Wu, and X.~Wu.
\newblock Achieving non-discrimination in data release.
\newblock In {\em Proceedings of the 23rd ACM SIGKDD International Conference
  on Knowledge Discovery and Data Mining}, pp. 1335--1344. ACM, Halifax, NS,
  Canada, 2017.

\bibitem{zhang2017causal}
L.~Zhang, Y.~Wu, and X.~Wu.
\newblock A causal framework for discovering and removing direct and indirect
  discrimination.
\newblock In {\em Proceedings of the 26th International Joint Conference on
  Artificial Intelligence}, pp. 3929--3935. AAAI Press, San Francisco, USA,
  2017.

\bibitem{zhang2017achieving}
L.~Zhang, Y.~Wu, and X.~Wu.
\newblock Achieving non-discrimination in prediction.
\newblock In {\em Proceedings of the Twenty-Seventh International Joint
  Conference on Artificial Intelligence, {IJCAI}}, pp. 3097--3103. {ijcai.org},
  Stockholm,Sweden, 2018.

\bibitem{Zhang2018examining}
Q.~Zhang, W.~Wang, and S.~Zhu.
\newblock Examining {CNN} representations with respect to dataset bias.
\newblock In {\em Proceedings of the Thirty-Second Conference on Artificial
  Intelligence}, pp. 4464--4473. AAAI press, New Orleans, Louisiana, USA, 2018.

\bibitem{vzliobaite2011handling}
I.~{\v{Z}}liobaite, F.~Kamiran, and T.~Calders.
\newblock Handling conditional discrimination.
\newblock In {\em 11th International Conference on Data Mining (ICDM)}, pp.
  992--1001. IEEE, Vancouver, BC, Canada, 2011.

\end{thebibliography}
